# Harnessing physical entropy noise in structurally metastable 1T' molybdenum ditelluride for true random number generation

Yang Liu[1,2], Pengyu Liu[1], Yingyi Wen[1], Zihan Liang[3], Songwei Liu[1], Lekai Song[1], Jingfang Pei[1], Xiaoyue Fan[4], Teng Ma[5], Gang Wang[4], Shuo Gao[6], Kong-Pang Pun[1], Xiaolong Chen[3], Guohua Hu[1,*]

[1]Department of Electronic Engineering, The Chinese University of Hong Kong, Shatin, N. T., Hong Kong S. A. R., China

[2]Shun Hing Institute of Advanced Engineering, The Chinese University of Hong Kong, Shatin, N. T., Hong Kong S. A. R., China

[3]Department of Electrical and Electronic Engineering, Southern University of Science and Technology, Shenzhen, 518055, China

[4]School of Physics, Beijing Institute of Technology, Haidian, Beijing, 100081, China

[5]Department of Applied Physics, Hong Kong Polytechnic University, Hung Hom, Kowloon, Hong Kong S. A. R., China

[6]School of Instrumentation and Optoelectronic Engineering, Beihang University, Beijing, 100191, China

*Correspondence to: ghhu@ee.cuhk.edu.hk

**KEYWORDS**: True random numbers, molybdenum ditelluride, ferroelectric polarization, physical entropy noise, cryptography.

**ABSTRACT**: True random numbers are essential in various research and engineering problems. Their generation depends upon a robust physical entropy noise. Here, we present true random number generation by harnessing the conductance noise probed in structurally metastable 1T' molybdenum ditelluride ($MoTe_2$). The noise, well-fitting a Poisson process, is proved a robust physical entropy noise at low and even cryogenic temperatures. Noise characteristic analysis suggests the noise may originate from the polarization variations of the underlying ferroelectric dipoles in 1T' $MoTe_2$. We demonstrate the noise allows for true random number generation, enabling their use as seed for generating high-throughput secure random numbers exceeding 1 Mbit/s, appealing for practical applications in, for instance, cryptography where data security is now a severe issue. As an example, we show biometric information safeguarding in neural networks by using the random numbers as mask, proving a promising data security measure in big data and artificial intelligence.

Random numbers, a string of random bits, play a crucial role in research and engineering, for example, serving as the random input in numerical simulations and modeling, and introducing uncertainty in gaming and decision-making. [1] However, the random numbers generated by deterministic algorithms are essentially pseudo-random. [2] As the demand for randomness is increasingly stringent, the need for true random numbers has become critical. [3] Physical entropy noise with inherent true randomness, such as thermal noise from thermally agitated charges and 1/f noise from charge trapping dynamics in semiconductors, can be employed for deriving true random numbers. [4] However, arising from their physical nature, these physical entropy noises can be vulnerable to ambient noise and cryogenic attacks, undermining their reliability for true random number generation. [5]

Nanomaterials with a plethora of quantum phenomena present promising solutions. [6–8] Amongst them, two-dimensional (2D) materials hold great interest, given their low-dimensionality and the quantum confinements, and the exhibited random variations in their underlying (opto)electronic and photonic processes. [9,10] They can also give flexible integration and synergy with modern electronics. [11,12] Indeed, reports show the random dynamics in emerging 2D materials and devices can be employed as robust physical entropy noises for true random number generators (TRNGs) and physically unclonable functions (PUFs) towards data and hardware cryptography. [13–18] 2D materials may even exhibit structural random variations. For example, molybdenum ditelluride ($MoTe_2$), a transition metal dichalcogenide, can exist in a prismatic-orthorhombic (2H-1T) intermediate octahedral 1T' phase. [19] The intermediate phase can induce metastable polarization of the underlying ferroelectric dipoles and, as a result, random variations in the electronic structures and properties. [20] Studies show the structural metastability, intrinsic to the 1T' phase, is even resilient to ambient noise and cryogenic attacks, [21,22] manifesting the potential of using 1T' $MoTe_2$ for true random number generation.

In this work, we report true random number generation from electrochemical-exfoliated 1T' $MoTe_2$. We show the 1T' $MoTe_2$ allows stable conductance noise probing and, notably, fitting a Poisson process, the noise is proved a robust physical entropy noise even at low temperatures down to 15 K. Noise characteristic analysis, including spectral density and statistical time-lag, suggests the noise arises from the polarization variations of the underlying ferroelectric dipoles in 1T' $MoTe_2$. Using the noise, we realize true random number generation, and prove high-throughput secure random numbers exceeding 1 Mbit/s, appealing for practical applications, such as cryptography. Here we show safeguarding neural networks by masking key biometric information with the random numbers, proving a promising data security measure for big data and artificial intelligence.

**Conductance noise probing**: $MoTe_2$ predominantly exists in a stable 2H phase (Fig. 1a). [23] The 2H phase, however, can have local lattice distortions along the $y$-axis, with the Te atoms forming an octahedral coordination around the Mo atoms, leading to phase transition to a structurally metastable 2H-1T octahedral ferroelectric 1T' phase. [23,24] 1T' $MoTe_2$ can be produced via physical and chemical engineering processes. [25,26] Towards scalable applications, here we adopt the electrochemical exfoliation method (Fig. 1b; Supplementary Note 1). [27] Briefly, tetrahexylammonium cations intercalate the bulk $MoTe_2$, leading to exfoliation and distortion of the 2H structure. Figure 1c presents a solution of the exfoliated $MoTe_2$. Transmission electron microscopic (TEM) images of the nanosheets (Fig. 1d, e) show the $MoTe_2$ is successfully exfoliated with minimal defects. Particularly, a lattice spacing of 3.4Å and a non-hexagonal structure are revealed, indicating the exfoliated $MoTe_2$ is in the 1T' phase. [28] Electron diffraction (Fig. 1e inset) shows a non-hexagonal rhombic, tetragonally symmetric lattice, confirming a 1T' phase. To verify the minimal defects, we perform X-ray photoelectron spectroscopic (XPS) analysis and prove a ~1:2 ratio for the Mo and Te atoms (Fig. S1). The minimal defects are ascribed to the tetrahexylammonium cation intercalation as the large molecules can effectively expand $MoTe_2$ while limiting the intercalations. [27] The minimal defects suggest the 1T' phase transition primarily arises from the intercalation-induced lattice distortion rather than defects.

The structural metastability in the 1T' phase may lead to random variations in the electrical properties. To probe the variations, we fabricate devices in a simple vertical structure, where the 1T' $MoTe_2$ is sandwiched between electrodes (Fig. S2; Supplementary Note 1). Indeed, as shown in Fig. 2a, the device shows resistance switching (switching ratio $>10^3$) with random variations under sweeping bias, proving a randomly varying electrical conductivity of the 1T' $MoTe_2$. The switching on the other hand suggests a crystalline structure distortion towards the metallic 1T phase [26] and/or a potential Stark modulation of the conductivity. [29,30] Nevertheless, the random variations may be harnessed as a physical entropy noise. For convenient operation, we measure the current output from the 1T' $MoTe_2$ devices under static bias instead of sweeping bias, and probe the random variations exhibited in the output.

We first study the output at 300 K and 0.05 V (Fig. 2b). See also Fig. S3a-h for the outputs at 0.1-5 V. As observed, stable variations are demonstrated in the outputs at all the bias conditions. For simplicity, we refer to the variations as conductance noise. We plot histograms of the outputs and fit Poisson process distributions, as shown in Fig. 2d (see also Fig. S3i-p). Note the outputs also conform to normal distributions (Fig. S4). Poisson process fitting is studied here to assess the randomness. A Poisson process describes a system of objects randomly distributed with an essential feature where the objects occur independently of one another. [31] As such, the Poisson process fitting proves the conductance noise is random, i.e. a physical entropy noise. [32] Given the request of the true random number applications for sustaining cryogenic attacks, we extend our assessment to cryogenic temperatures (Fig. 2c). As shown, stable conductance noise is proved at all the low temperatures, and the outputs can fit Poisson processes (Fig. 2e-g). Here we note the Poisson fittings at the varying temperatures develop essentially consistent fitting characteristics, and that a larger $\lambda$ at the higher temperatures means a larger averaged number of occurring events per interval only. [31] Nevertheless, our investigation proves the conductance noise is a robust physical entropy noise stable even at cryogenic temperatures.

As discussed, noise from thermal and ambient electronics stands as a key source of physical entropy noise. [4] We concurrently test ambient noise to study whether the conductance noise is due to the ambient noise from thermal and/or the electrical test set-ups. As shown in Fig. 2b, c, the ambient noise is smaller than the conductance noise by several orders of magnitude, proving the conductance noise is primarily from 1T' $MoTe_2$ rather than the ambient noise. We also test the device-to-device and batch-to-batch conductance noises (Fig. S5 and S6). From the tests, we learn that the device fabrication approach is reliable, with an acceptable yield in the initial trials (~70-80%), and that the measured conductance noises all can well fit Poisson processes, though the outputs may be different in the current values. Further materials and device engineering are required to improve the device fabrication yield and uniformity.

**Origin of the conductance noise**: To study the origin of the conductance noise, we perform current noise power spectral density (PSD) testing of the 1T' $MoTe_2$ devices at different bias and temperature conditions (Fig. 3a-d, Fig. S7). As shown, 1/f noise is proved over a broad frequency. For example, at 0.05 V, 300 K, the noise spectral power well fits 1, i.e. $\gamma \sim 1$, proving 1/f noise (Fig. 3c). However, as the frequency increases, the output may be flattened at low bias conditions, indicating thermal noise dominates the high-frequency region (Fig. 3d). The flattening may be ascribed to the insufficient current signals at low bias conditions. [33,34] Nevertheless, 1/f noise is proved for our 1T' $MoTe_2$ devices, suggesting charge fluctuations account for the conductance noise, for instance, charge trapping dynamics at the defect sites [35,36] in our 1T' $MoTe_2$. However, as observed, varying the bias and temperature does not really variate the 1/f noise, suggesting charge trapping by defects may not account for the conductance noise. Indeed, as discussed, our 1T' $MoTe_2$ have minimal defects. We therefore understand the conductance noise arises from the intrinsic property of 1T' $MoTe_2$ rather than the defects. Given the structural metastability of the ferroelectric 1T' phase, we assume the conductance noise is the reflection of the polarization variations of 1T' $MoTe_2$. However, ferroelectric polarization (Fig. S8) and Raman spectroscopy (Fig. S9) characterizations show no deterministic collective, overall ferroelectric polarization. The collective polarization may have been compromised by the discrete nature of the exfoliated 1T' $MoTe_2$ nanosheets [37], and further characterizations on the individual nanosheets will be required to locate the exact polarization effect of the underlying ferroelectric dipoles. With the above investigations, we understand the conductance noise arises from the polarization variations of the ferroelectric dipoles in the individual 1T' $MoTe_2$ nanosheets rather than a collective polarization.

To investigate the origin further, we study the cumulative charge characteristic of the devices. Here we plot in Fig. 3e the output at 0.05 V, 300 K and the cumulative charge integrated during the sampling time intervals. The cumulative charge state can reflect the polarization state of the ferroelectric dipoles in 1T' $MoTe_2$. [38] As observed, the cumulative charge proves a stable noise, indicating random variations in the polarization states. We adopt weighted time-lag, a method widely used to study noise characteristics, [39] to statistically evaluate this noise. Briefly, as plotted in Fig. 3f, the distribution of the cumulative charge is defined with a weighted time-lag, $TL =$ $lg\left(K\sum_{n}^{N-1}\frac{1}{2\pi\alpha^2}\exp\left(-\frac{[(Q_n-x)^2+(Q_{n+1}-y)^2]}{2\alpha^2}\right)\right)$, where $Q_n$ and $Q_{n+1}$ are the $n$-th and $(n+1)$-th cumulative charge states, $(x, y)$ denotes the corresponding $TL$ plot coordinates, $N$ is the total states, and $\alpha$ and $K$ are the fitting parameters to ensure the $TL$ maximum is 1 before logarithm. The cumulative charge states in the plot are distributed in ascending order, and a $TL$ approaching 0 means a stronger correlation between the adjacent states. As observed, the $TL$ plot shows random yet uniform aggregations with a bimodal distribution along the diagonal – both the larger and smaller cumulative charge states establish stronger correlations, while the medium cumulative charge states weaker correlations. This indicates the ferroelectric dipoles are uniformly distributed with bimodal aggregations, and that the strongly polarized dipoles may require a stronger current and a longer time to reverse.

See Fig. S10 for the control experiments on the ambient noise and a $MoS_2$ device, where we demonstrate the ambient noise and random telegraph noise develop distinctly different cumulative charge profiles and $TL$ patterns. The comparison aligns with our understanding that the conductance noise arises from the polarization variations of the underlying ferroelectric dipoles in 1T' $MoTe_2$.

We conduct Monte Carlo simulation on the polarization states of the ferroelectric dipoles in 1T' $MoTe_2$ (as illustrated in Fig. 3g). See Supplementary Note 2 for the simulation method. Based on our understanding, the ferroelectric dipoles are uniformly distributed and undergo polarization variations. Under a static bias, the polarization variations can lead to fluctuations of the bound charges and thus, the conductance noise. As studied by the Monte Carlo simulation, the output renders a stable noise, consistent with our experimental results (Fig. 3h); the conductance noise also well fits a Poisson process (Fig. 3i). Therefore, the Monte Carlo simulation from the perspective of bound charge fluctuations aligns with our understanding that the conductance noise arises from the polarization variations of the ferroelectric dipoles.

**True random number generation**: We design a very simple circuit in Cadence Virtuoso to harness the conductance noise for true random number generation (Fig. 4a, b). The circuit consists of a 1T' $MoTe_2$ device, an I/V converter, a high pass filter, a voltage amplifier, and a comparator. Upon operation, the converter transforms the output from the 1T' $MoTe_2$ device into a voltage signal for convenient signal processing. The voltage signal, i.e. '*Output 1*' (Fig. 4c), demonstrates a noise that can well fit a Poisson process (Fig. 4d). This shows the circuit has well captured the conductance noise from the 1T' $MoTe_2$ device. The voltage signal then passes through the filter and amplifier to extract the noise in the form of differentiated voltage spikes, i.e. '*Output 2*' and '*Output 3*' (Fig. 4c). As demonstrated, the voltage spikes also fit a Poisson process (Fig. 4e, f), proving the extracted noise remains a robust physical entropy noise. The comparator then processes the voltage spikes to yield random numbers in the form of 0's and 1's binary strings, i.e. '*Output 4*' (Fig. 4c). See also Fig. S11 for a zoomed-in distribution of the 0's and 1's digits. As demonstrated, the random 0's and 1's digits render a ratio of ~1:1 (Fig. 4g), suggesting the numbers are random.

To verify the random numbers are truly random, we test the randomness using the National Institute of Standards and Technology (NIST) randomness testing suite that is widely used to evaluate the randomness. [40] As shown in Table S1, the random numbers successfully pass the test, proving their true randomness. However, the throughput is ~10 bit/s, limiting their use.

To address the limitation, employing a common approach in the field, [41] we use the random numbers as seed and introduce them to a Non-Linear Feedback Shift Register (NLFSR; Fig. S12) for high-throughput random number output with a rate of, say, 1 Mbit/s (Fig. 4a, b). Here, as a case example, we present in Fig. S13 a random number bitmap generated using the high-throughput random numbers. Notably, as shown in Table S2, the high-throughput random numbers also fully pass the NIST randomness test.

Here we note the electrical components (e.g. the resistors and comparators) in Fig. 4b are configured to adapt to room temperature operation. For low temperature (e.g. 15 K) operations, the electrical components need to be reconfigured to adapt to low current values, and special temperature-compensating and corrective circuits may be used. [42] Nevertheless, the demonstrated capability of our 1T' $MoTe_2$ devices for surviving cryptogenic attacks is critical for TRNG applications, outperforming the current TRNG techniques (Table S3).

**Safeguarding neural networks**: High-throughput secure random numbers are appealing for practical applications. [3] As an example, we apply the high-throughput secure random numbers in cryptography. Cryptography is of critical importance in the current age with the data exponentially growing and at the risk of being attacked and sabotaged. [3] See Fig. S14 and S15 for demonstrations of password generation and data encryption.

Beyond the common cryptographic applications, the importance of cryptography is manifested by the rapid advances of neural networks. Neural networks are widely used in, for instance, image recognition, sensing, autonomous driving, and manufacturing, where sensitive data is constantly required. Particularly, in the context of big data and artificial intelligence, sensitive biometric information such as retinas, facial characteristics, and fingerprints is excessively used and potentially leaked. [43] Secure data protection has therefore become a critical issue, and adversarial attacks in neural networks pose a significant threat. [43] Here we adopt a differential safeguarding strategy (Fig. S16) and investigate its effectiveness to obfuscate sensitive data in neural networks. [43] The differential safeguarding framework injects random numbers as noise into the target data for perturbation. Following this approach, we train a residual neural network (ResNet) for pet recognition (Fig. 5a; Supplementary Note 3). ResNets are a widely used model for image and pattern recognition. [44] After training, the model performs successful pet recognition with an accuracy of 92%. See the confusion matrix and performance of the well-trained model in Fig. S17.

We then inject the random numbers as noise into the target validation data. Interestingly, as shown in Fig. 5a, taking the image of a Siamese cat for demonstration, the noise perturbation appears negligible to human eyes. This is ascribed to the innate ability of the human brain to process visual information holistically, i.e. focusing on the broader picture rather than the minute details. [45] However, as demonstrated in Fig. 5b, the noise perturbation substantially affects the recognition of the model at all the convolution layers. The confusion matrix and accuracy (~78%) demonstrate the ResNet variant with the noise perturbation gives a poor performance in recognition (Fig. 5c, d). See also Fig. S18 for the confusion matrix details with noise perturbation. Comparing the accuracies with and without the noise perturbation (Fig. 5d), a bit noise perturbation can cause a substantial degradation in the classification accuracy. This is because the noise disrupts the feature detection capability of the ResNet variant model in the initial layers, which then propagates to the deeper layers, leading to exacerbation of the errors. [46] The detailed difference (i.e. Δ) between the two success rates in the different 37 categories can be found in Fig. 5e. The findings prove that injecting the random numbers as noise perturbations that are not discernible to eyes can substantially interfere with the neural networks and as such, enhance data security in neural networks.

In conclusion, we have reported true random number generation using structurally metastable 1T' $MoTe_2$. Our analysis suggests the polarization variations of the ferroelectric dipoles in 1T' $MoTe_2$ give rise to a conductance noise, and that the noise can be harnessed as a robust physical entropy noise for true random number generation. Notably, the noise remains stable even at cryptogenic temperatures, critical for the cryptogenic applications of true random numbers. We have proved true random number generation and their use as the seed for high-throughput secure random number generation exceeding 1 Mbit/s, appealing for various practical applications, for instance, cryptography. As an example, we demonstrate data safeguarding in neural networks by using the random numbers as the mask. Neural networks pose a significant threat to data security, particularly the biometric information, in big data and artificial intelligence by adversarial data leakage and attacks. Our safeguarding approach can serve as a critical data security measure. Given this, and the scalability as well as seamless integration possibility with the electronics systems of the electrochemical-exfoliated 1T' $MoTe_2$, our approach of true random number generation holds great potential to enable secure data in neural networks.

**FIGURES**

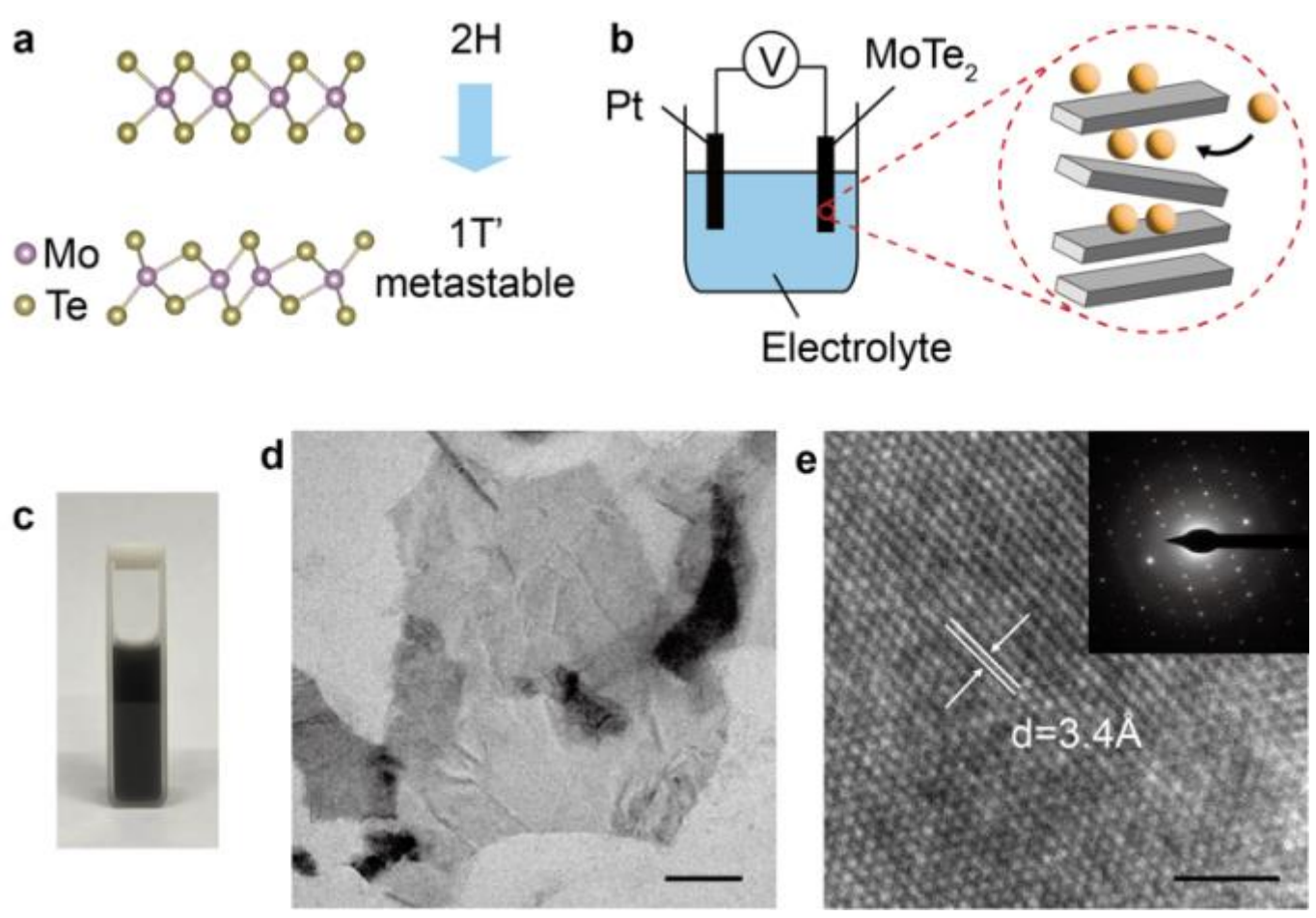

**Figure 1. 1T' $MoTe_2$ by electrochemical exfoliation.** (a) Crystalline structures of hexagonal 2H and distorted octahedral 1T' $MoTe_2$. (b) Schematic electrochemical exfoliation of $MoTe_2$, showing intercalation of molecular tetrahexylammonium cations between the $MoTe_2$ layers. Pt and $MoTe_2$ are used as the electrodes. (c) Dispersion of the exfoliated $MoTe_2$. (d, e) Transmission electron microscopic images and the selected electron diffraction pattern (inset of e) of the exfoliated $MoTe_2$ nanosheets, proving a distorted octahedral 1T' crystalline structure.

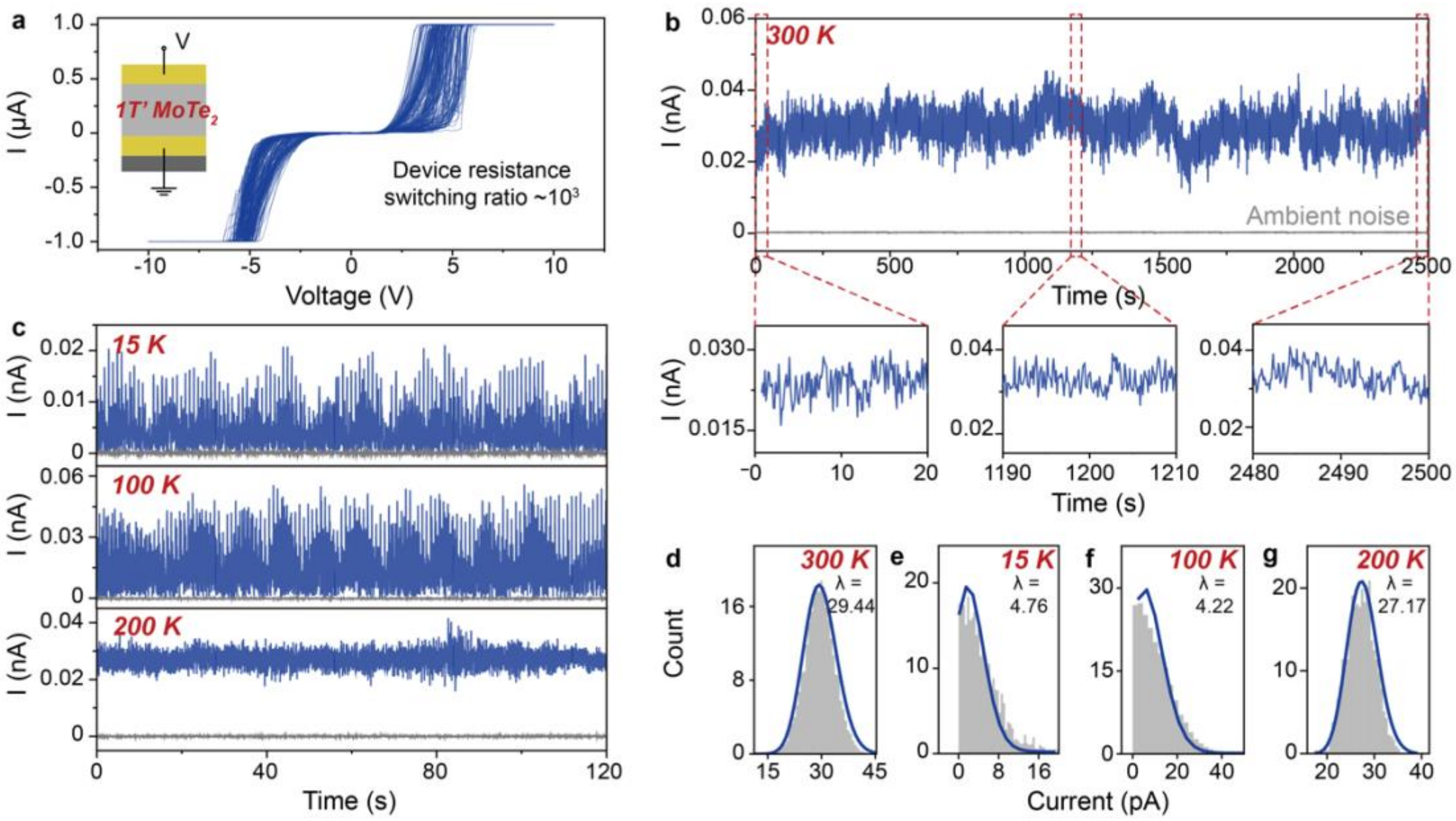

**Figure 2. Conductance noise probing.** (a) Current output for 100 sweeping test cycles of a typical 1T' $MoTe_2$ device, showing random variations in the device resistance switching at ~5 V. The switching ratio is estimated as 1424 between the averaged high and low resistances. The inset shows the device structure. (b) Current output from the device at 300 K, showing stable conductance noise probed from 1T' $MoTe_2$. (c) Current output at low temperatures, showing stable conductance noise. The bias is 0.05 V for all the tests. See Fig. S3 for the output at the other bias conditions. Ambient noise measured from suspended probing electrodes is presented in grey for comparison. (d-g) Histograms and Poisson fittings of the current output from (b) and (c), proving the conductance noise probed is a random process.

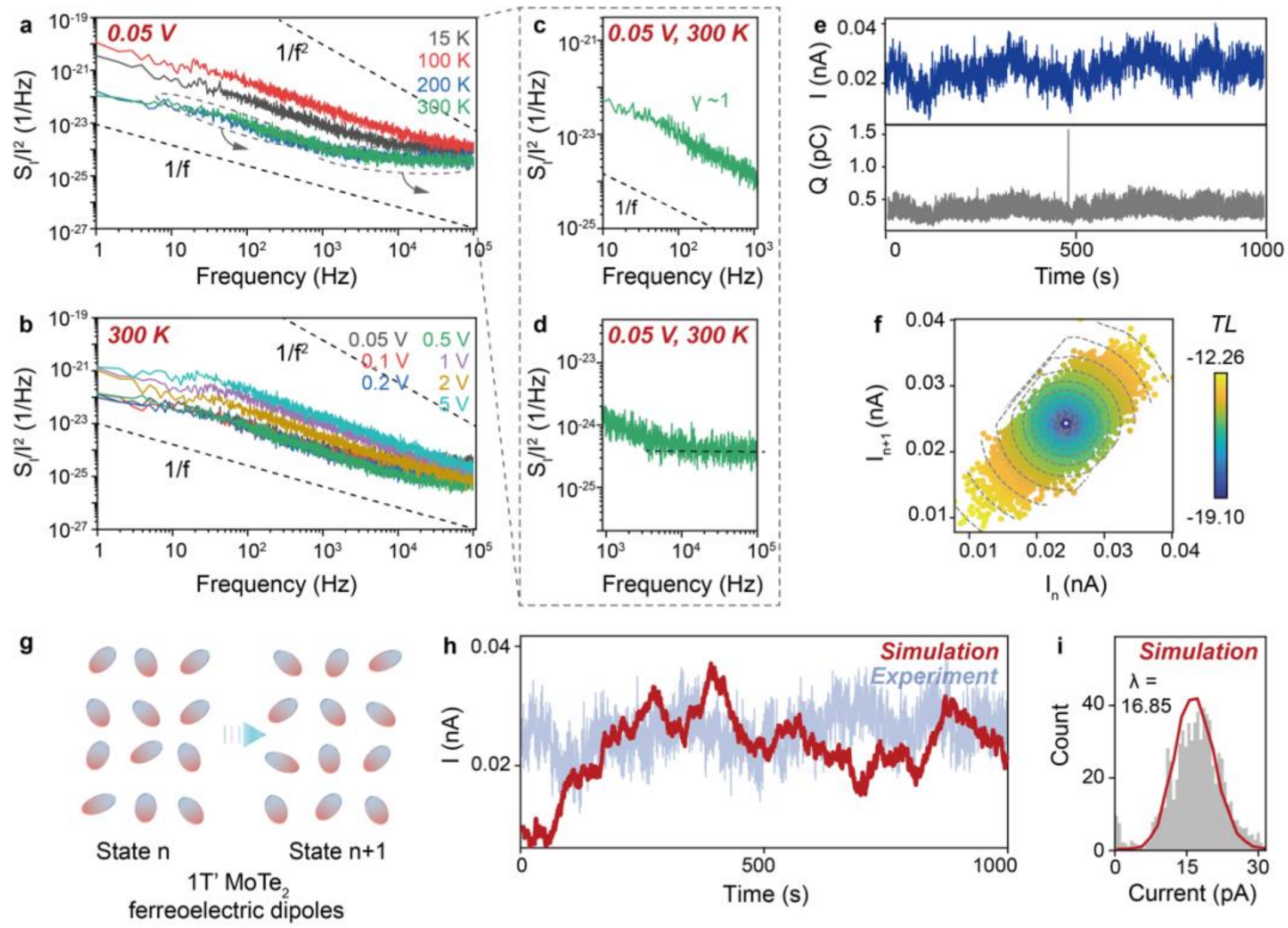

**Figure 3. Origin of the conductance noise.** Current power spectral density (PSD) of the 1T' $MoTe_2$ device at (a) 0.05 V and (b) 300 K, with (c, d) plotting PSD at low and high-frequency regions. PSD testing proves 1/f noise. (e) Current output and the corresponding cumulative charge fluctuation of the 1T' $MoTe_2$ device at 0.05 V, 300 K. The cumulative charge is integrated in the sampling interval of 0.067 s. (f) Time-lag plot for the cumulative charge fluctuation, showing bimodal distribution along the diagonal, suggesting stronger collations for the larger and smaller cumulative charge states to the next states. (g) Varying $n$-th and $(n + 1)$-th polarization states of the ferroelectric dipoles in 1T' $MoTe_2$ in our proposed conductance noise mechanism. (h) Current output from Monte Carlo simulation on polarization variations of the ferroelectric dipoles, with the current output in (e) plotted as the background for comparison. (i) Histogram and Poisson fitting of the current output from Monte Carlo simulation.

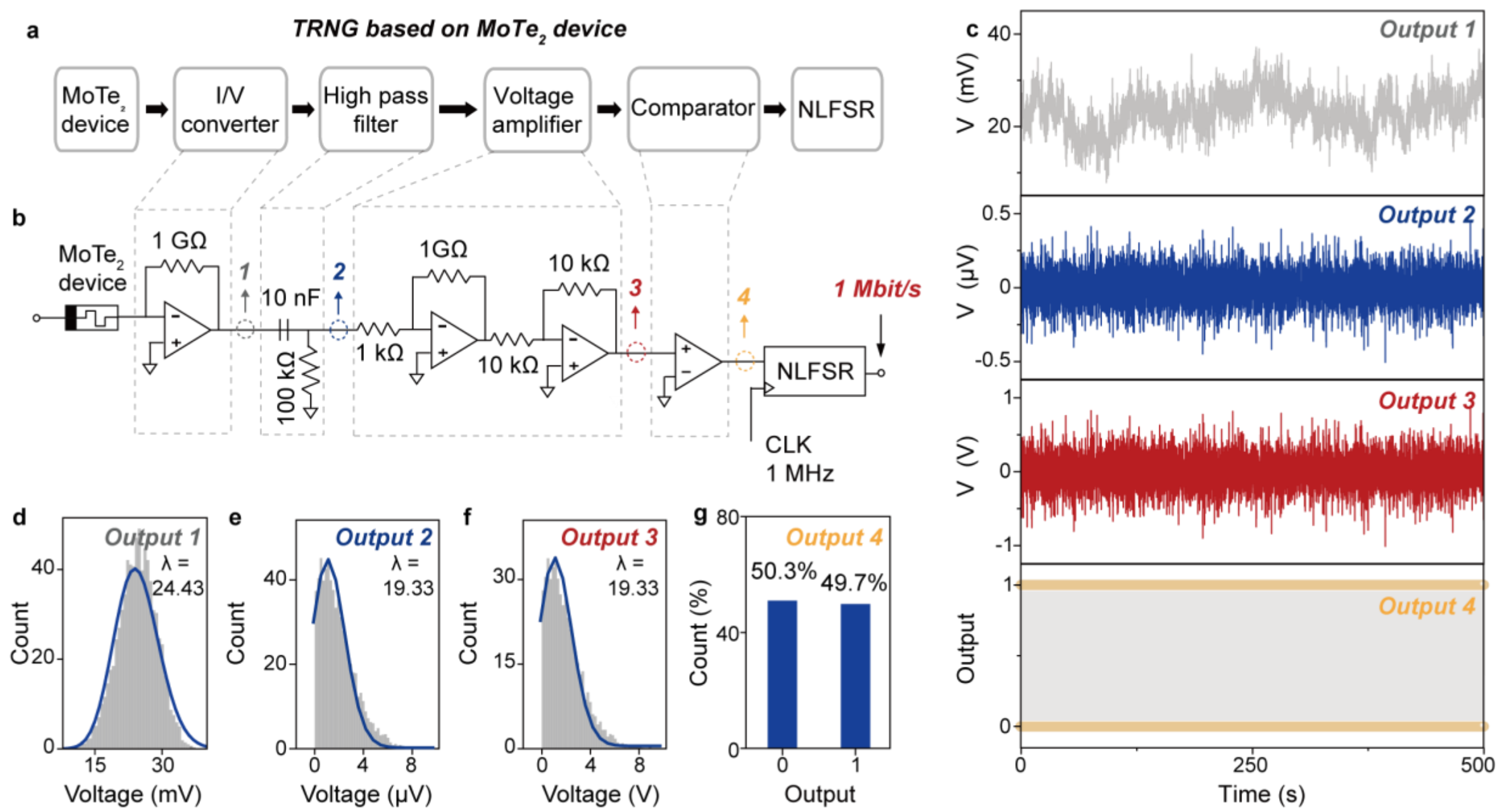

**Figure 4. True random number generation.** (a, b) Circuit design and diagram of the true random number generator (TRNG), consisting of the 1T' $MoTe_2$ device, I/V converter, high pass filter, voltage amplifier, comparator, and NLFSR. NLFSR (Fig. S12) is used to generate high-throughput random numbers from the seed, i.e. the true random numbers from Port 4. (c) The output obtained from the TRNG at Port 1-4, including the converted voltage output, the filtered voltage output, the amplified voltage output, and the generated true random numbers in a string of 0's and 1's. See Fig. S11 for the 0's and 1's string details. (d-f) Histograms and Poisson fittings of the data points from Output 1-3 in (c), proving all the outputs are random processes. Note absolute values are taken from the negative data points in Output 2 and 3 for the Poisson fittings. (g) Histogram of the 0's and 1's true random numbers in Output 4 in (c), demonstrating a ratio of ~1:1.

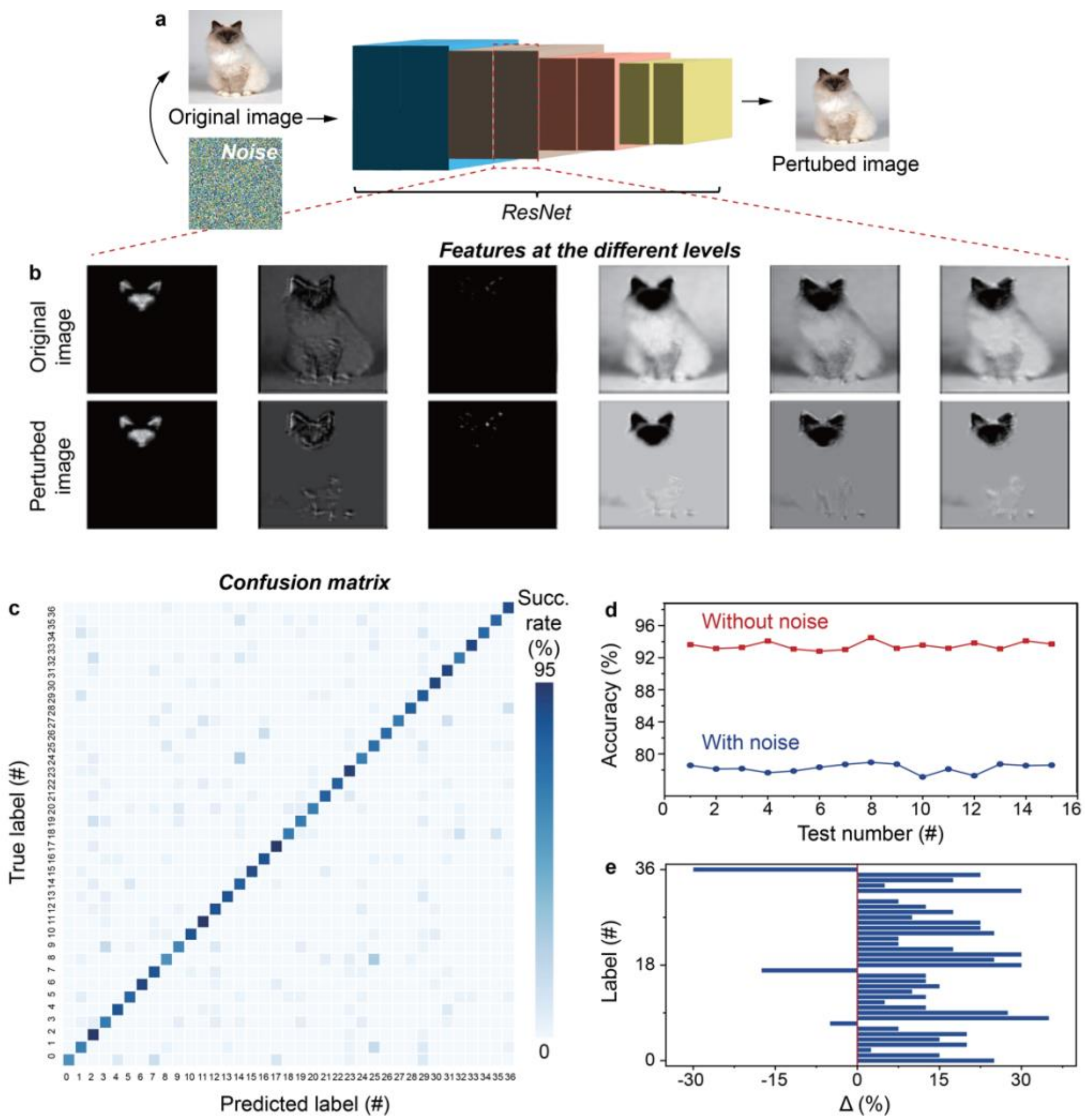

**Figure 5. Safeguarding in neural networks.** (a) ResNet variant architecture for cat recognition without and with noise perturbation. The noise map is produced using the high-throughput random numbers. (b) The cat images without and with noise perturbation at the intermediate convolution layers, showing the perturbed images lose certain features of the cat throughout the convolution layers. (c) Confusion matrix for the ResNet variant recognition with noise perturbation. The scale corresponds to the success rate of the predicted labels. The $x$ and $y$ coordinates denote the predicated and true labels of the 37 classifications in the training dataset. (d) The accuracy with and without noise perturbation at the different test numbers. (e) The difference between the success rates in the confusion matrices with and without noise perturbation along the diagonal. The $x$

coordinate denotes the difference in the success rate along the diagonal, and the $y$ coordinate the 37 different classifications. The confusion matrix details with the success rate values are presented in Fig. S18.

## Author Contributions

YL, GH designed the experiments. YL, PL, YW, ZL, SL, LS, JP, XF performed the experiments. YL, PL, GH analyzed the data. YL, GH prepared the figures. YL, GH wrote the manuscript. All authors discussed the results from the experiments and commented on the manuscript.

## Funding Sources

GH acknowledges support from CUHK (4055115), YL from SHIAE (RNE-p3-21), JP and YW from RGC (24200521), TM from PolyU (P0042991), GW from NSFC (12074033), SG from National Key Research and Development Program of China (2023YFB3208003), XC from NSFC (62275117) and Shenzhen excellent youth program (RCYX20221008092900001).

**Supplementary Notes**

**Supplementary Note 1: Material production and device fabrication**

The raw $MoTe_2$ powder and other chemicals are purchased from Alpha Aeser and Sigma-Aldrich, and are used as received. The Raw $MoTe_2$ powder is pressed with high pressure into a solid disk for the electrochemical exfoliation. The electrochemical exfoliation follows the method reported in Ref. (1). For device fabrication, the Au/$MoTe_2$/Au devices are fabricated on Si/$SiO_2$, where $MoTe_2$ is deposited by spin-coating, and the gold electrodes are deposited by electron-beam evaporation. The electron-beam evaporator is IVS EB-600. The $MoTe_2$ after deposition is baked at 400 $^oC$ for 0.5 hours under nitrogen.

Tektronix Keithley 4200-SCS parameter analyzer is used to measure the electrical characteristics of the devices under 300 K at ambient conditions. For the 15 K, 100 K, and 200 K tests, FS-Pro is used under vacuum (~$10^{-6}$ mbar).

**Supplementary Note 2: Monte Carlo simulation**

Assuming a constant electric field, the change in the polarization of the ferroelectric dipoles will not affect the field, but the internal polarization state of the ferroelectric dipoles in the 1T' $MoTe_2$ will change. The fluctuation of the polarization can lead to fluctuating bound charges, which can in turn cause fluctuations in the conductance state of the 1T' $MoTe_2$.

The polarization can switch between two states, i.e. $P_1$ and $P_2$, and the switching follows the Arrhenius law [2]. The polarization can thus be modelled as a Poisson process, meaning that the probability of a switch in a small interval of time $dt$ is given by λ*dt, where λ is the rate of the process. The rate follows the Arrhenius law, given by $\lambda = \mathrm{A} * \exp(-\mathrm{E}/(\mathrm{k} * \mathrm{T}))$, where $A$ is the pre-exponential factor, $E$ is the energy barrier, $k$ is the Boltzmann constant, and $T$ is the temperature. However, the switch in polarization now results in fluctuations in the bound charge instead of a change in the electric field. The bound charge $\rho_B$ is related to the polarization $P$ by the relation $\rho_B = -divP$, where $div$ is the divergence operator, indicating that the bound charge is related to the spatial variation of the polarization. In a simple one-dimensional case, this can be described as $\rho_B = -dP/dx$. Now assume that the change in polarization is uniform across the material, the bound charge will change by $\Delta\rho_B$ proportional to the change in polarization ΔP, expressed as $\Delta\rho_B = -\Delta\mathrm{P}/\mathrm{L}$, where $L$ is a characteristic length scale of the system. This change in voltage can cause a current to flow.

We propose that the change in the bound charge affects the resistance $R$ of the material. A simple model assumes that the resistance is inversely proportional to the absolute value of the bound charge, $\mathrm{R} = \mathrm{R}_0/\rho_B$, where $\mathrm{R}_0$ is a constant initial resistance. Finally, apply a constant voltage $V$ across this variable resistor, the current $I$ through the material at any time would be given by Ohm's law, $I(t) = V/R(t)$. The charge from the current can be described by $Q_i = \int_{t=i}^{t=i+1} dI/dt$. So every time the polarization switches, it will change the bound charge, which will then change the resistance and hence, the current fluctuations.

Consider a common effect often found in materials known as Poole-Frenkel behavior [3], the current through the material (and hence the resistance) is affected by the applied electric field (which in our case can be linked to the bound charge), and the current density J is given by $J = J_0 * \exp(\beta * \sqrt{\varepsilon})$, where $J_0$ is the current density at zero field, $\varepsilon$ is the electric field, and β is a material constant. In our scenario, we now link the bound charge and the electric field $\varepsilon$. Assume that the change in the bound charge $\Delta\rho_B$ is proportional to the change in electric field $\Delta\varepsilon$, the bound charge is expressed by $\Delta\rho_B = -\Delta\varepsilon/L$. Then, the current density will be dependent on the bound charge. Note that the $sqrt(\varepsilon)$ means that this is not a simple linear or inversely proportional relationship.

## Supplementary Note 3: Neural network recognition

The neural network security is carried out in Python 3 and is based on the ResNet framework. The ResNet variant is based on a ResNet 34 structure, consisting of convolution layers, residual blocks, and so on. The information about the ResNet 34 can be found at: *https://pytorch.org/vision/main/models/generated/torchvision.models.resnet34.html*. The public dataset is from the Visual Geometry Group at the University of Oxford (available at *https://www.robots.ox.ac.uk/~vgg/data/pets/*). The dataset consists of a 37-category pet dataset with roughly 200 images for each class with different scales, poses, and lighting. All the images have an associated ground truth annotation of breed, head ROI, and pixel-level trimap segmentation, and those are used for training and testing.

**Supplementary Figures**

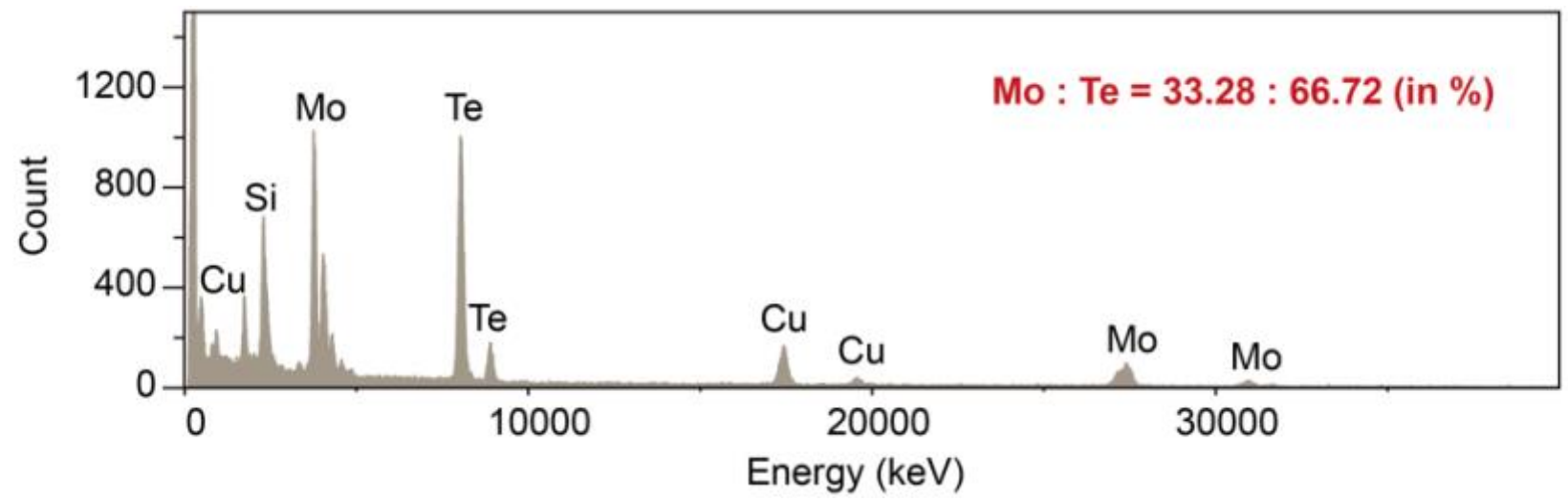

**Figure S1. Energy dispersive X-ray spectrometry element mapping of the exfoliated 1T' $MoTe_2$ nanosheets.** The ratio of the Mo and Te atoms is 33.28: 66.72 (in %), proving that there are minimal defects in the exfoliated 1T' $MoTe_2$ nanosheets.

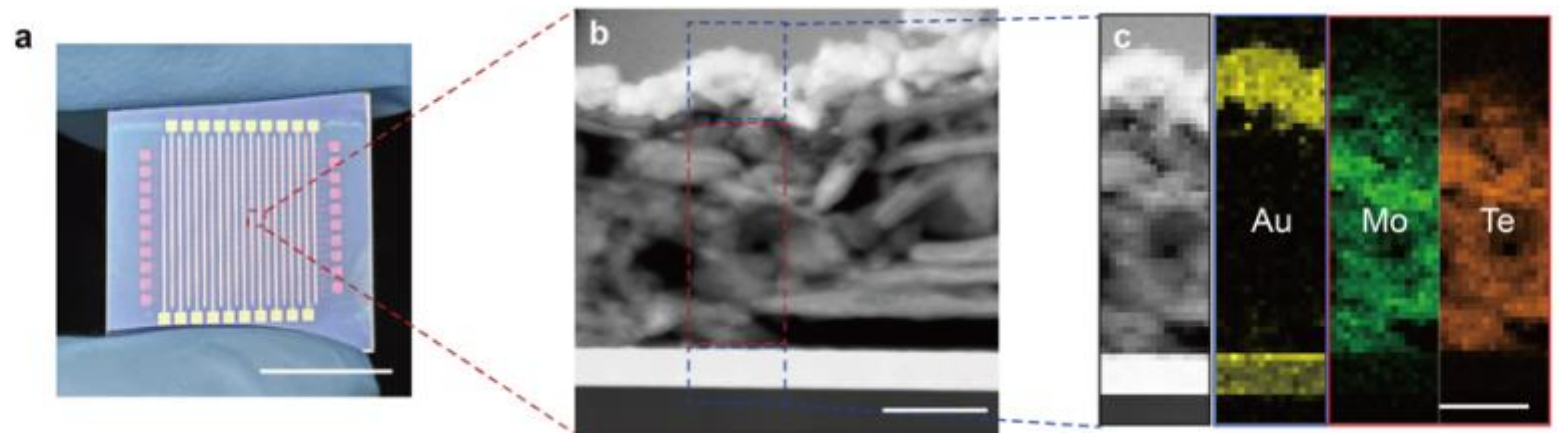

**Figure S2. 1T' $MoTe_2$ device array.** (a) Image of an array of 20 ×20 1T' $MoTe_2$ devices. Spin-coated 1T' $MoTe_2$ is sandwiched between the evaporated top and bottom gold electrodes to fabricate the devices. The substrate is $Si/SiO_2$. (b) Cross-sectional scanning electron microscopic image of a typical device, and (c) the corresponding elemental analysis of the selected areas for the Au, Mo, and Te elements, showing clear interfaces between the $MoTe_2$ layer and the electrodes. Scale bars – (a) 1 cm, (b) 300 nm, and (c) 300 nm.

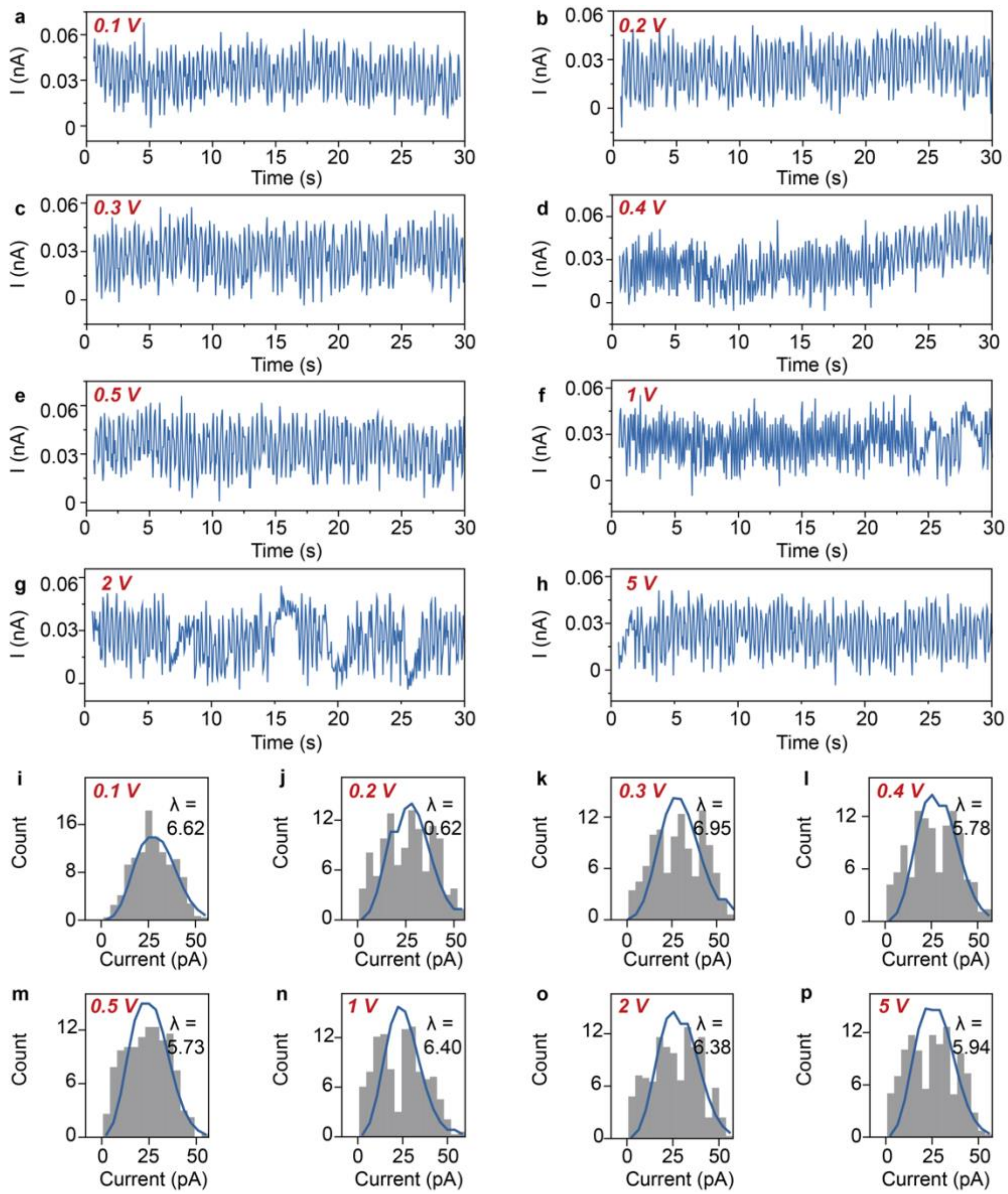

**Figure S3. Conductance noise probing at different biases.** (a)-(h) Current output of a 1T' $MoTe_2$ device at bias of 0.1 V, 0.2 V, 0.3 V, 0.4 V, 0.5 V, 1 V, 2 V, and 5 V. The temperature is 300 K. (i)-(p) The corresponding histograms and Poisson fittings of the current data points, proving the conductance noise in the 1T' $MoTe_2$ device measured at all the bias conditions is a random process and a reliable physical entropy noise.

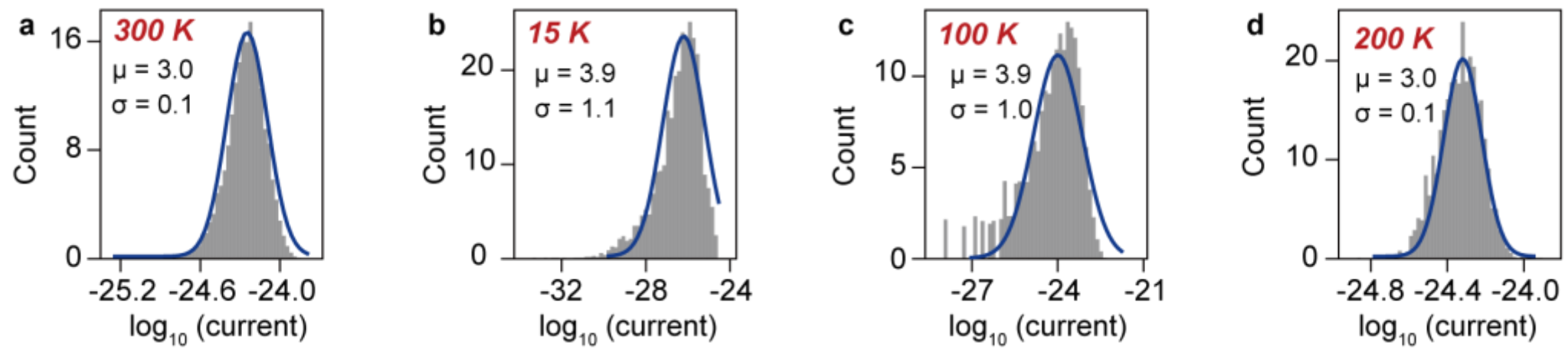

**Figure S4**. **Semi-log-normal distribution of the 1T' $MoTe_2$ device current output.** Histograms and semi-log-normal distribution fittings of the current outputs from Fig. 2b and c.

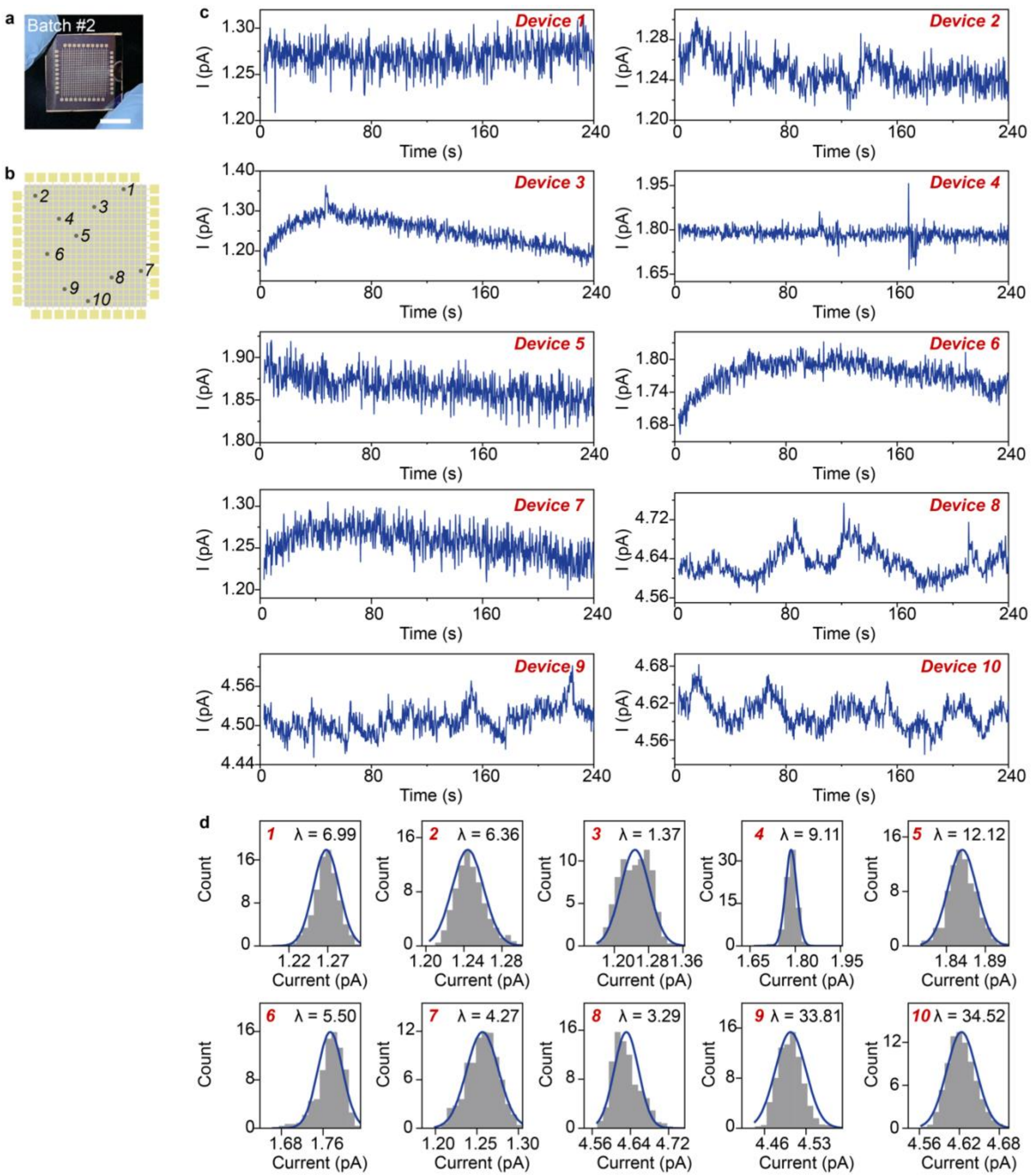

**Figure S5. Batch #2 device-to-device conductance noise.** (a) 20 ×20 1T' $MoTe_2$ devices, with a fabrication yield of ~70%, and (b) the corresponding schematic device array showing the working devices randomly selected for the sampling test. (c) Current noise as measured from the selected devices, and (d) the corresponding histograms and Poisson fittings of the current data from (c). Scale bar – (a) 1 cm**.**

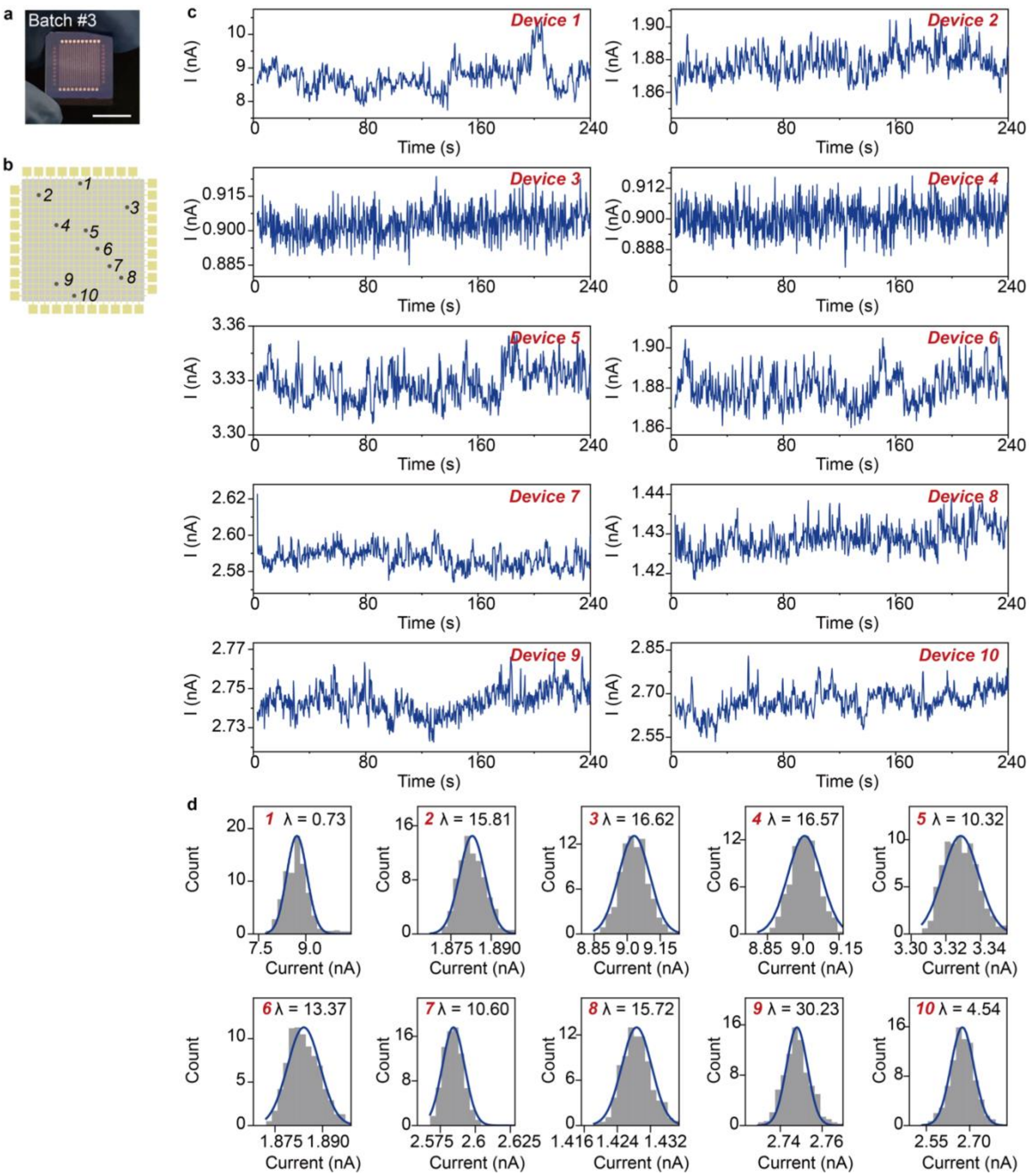

**Figure S6. Batch #3 device-to-device conductance noise.** (a) 20 ×20 1T' $MoTe_2$ devices, with a fabrication yield of ~80%, and (b) the corresponding schematic device array showing the working devices randomly selected for the sampling test. (c) Current noise as measured from the selected devices, and (d) the corresponding histograms and Poisson fittings of the current data from (c). Scale bar – (a) 1 cm.

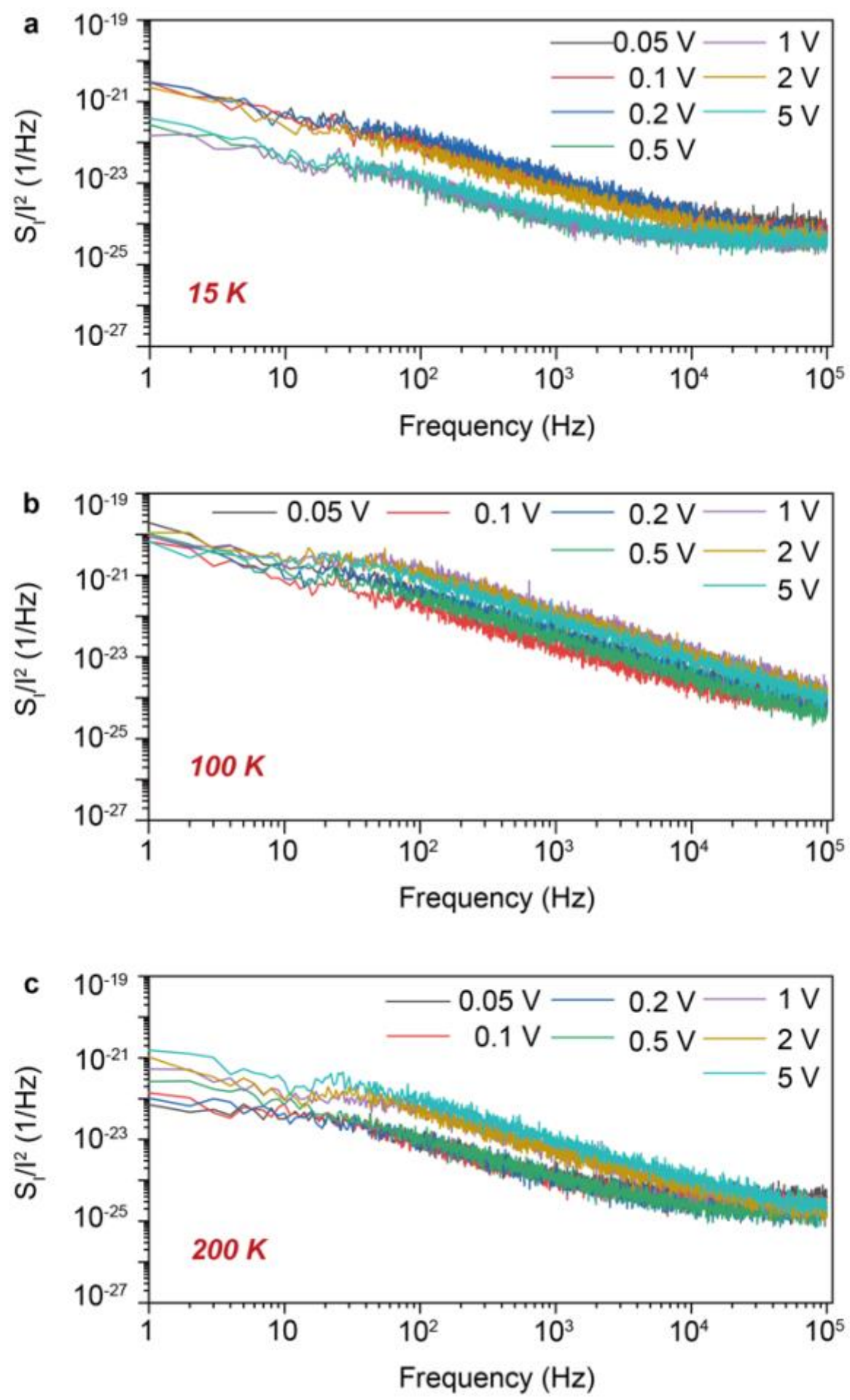

**Figure S7. Current power spectral density (PSD) characterization of the 1T' $MoTe_2$ device at low temperatures.** The PSD testing is conducted on the 1T' $MoTe_2$ device at a low temperature from 15 K to 200 K at the different bias conditions – (a) 15 K, (b) 100 K, and (c) 200 K. The testing proves 1/f noise in the device for all the temperatures and bias testing conditions in the low-frequency region.

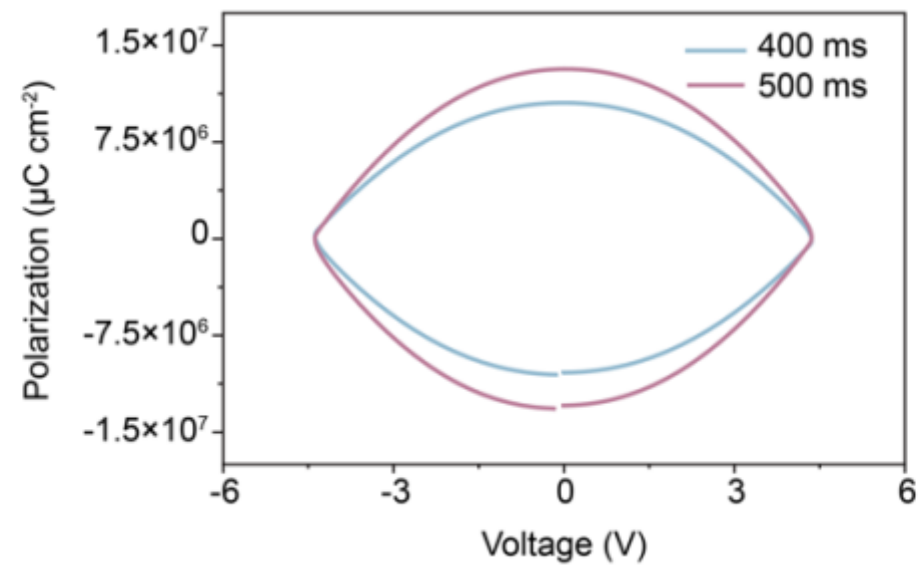

**Figure S8. Ferroelectric polarization characterization of the 1T' $MoTe_2$ device.** The ferroelectric polarization testing is conducted on a typical 1T' $MoTe_2$ device with a sweeping bias, showing no clear evidence of ferroelectric polarization. The test condition: $V_{max}$ = 5 V; Hysteresis speed = 400 ms/500 ms; Preset delay = 1,000 ms. No significant collective polarization is observed in this testing. Further ferroelectric polarization mapping over individual 1T' $MoTe_2$ nanosheets is required to deterministically locate the exact polarization effect of the ferroelectric dipoles.

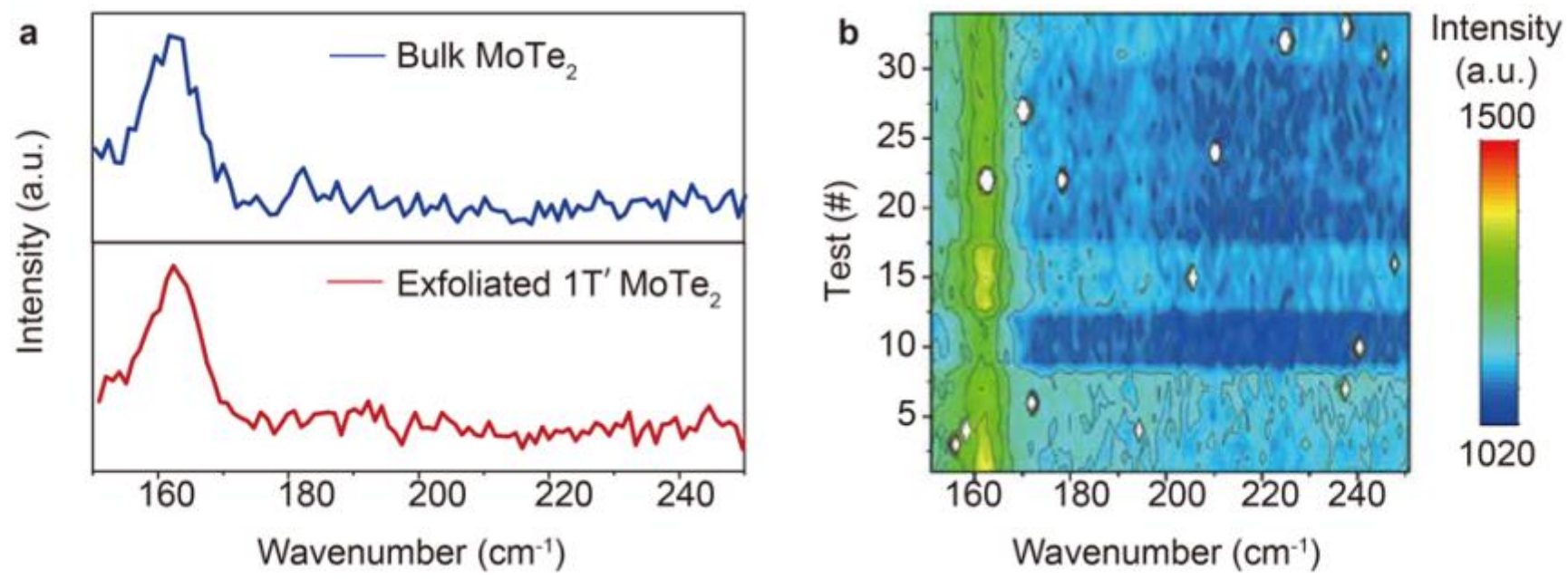

**Figure S9. Raman spectroscopy characterization of 1T' $MoTe_2$.** (a) Raman spectrum of bulk and the exfoliated 1T' $MoTe_2$. (b) Raman spectrum for the same area taken for 35 times. The test condition: wavelength 532 nm, power 2 mW, integration time for a single test 300s, and spot diameter size 1.5 um. No deterministic Raman peak shifts can be observed. Two main reasons may account for this – 1) as the spot size is 1.5 um in diameter for the Raman spectroscopy characterization, the measured Raman spectrum for the exfoliated 1T' $MoTe_2$ nanosheets is a collective, averaged Raman spectrum of many 1T' $MoTe_2$ nanosheets; and 2) each of the Raman spectroscopy tests takes 300 s, and this long time duration is not able to present the variations of the metastable phase structure and ferroelectric polarization. Further Raman spectroscopy on individual 1T' $MoTe_2$ nanosheets is required to deterministically locate the exact polarization effect of the ferroelectric dipoles.

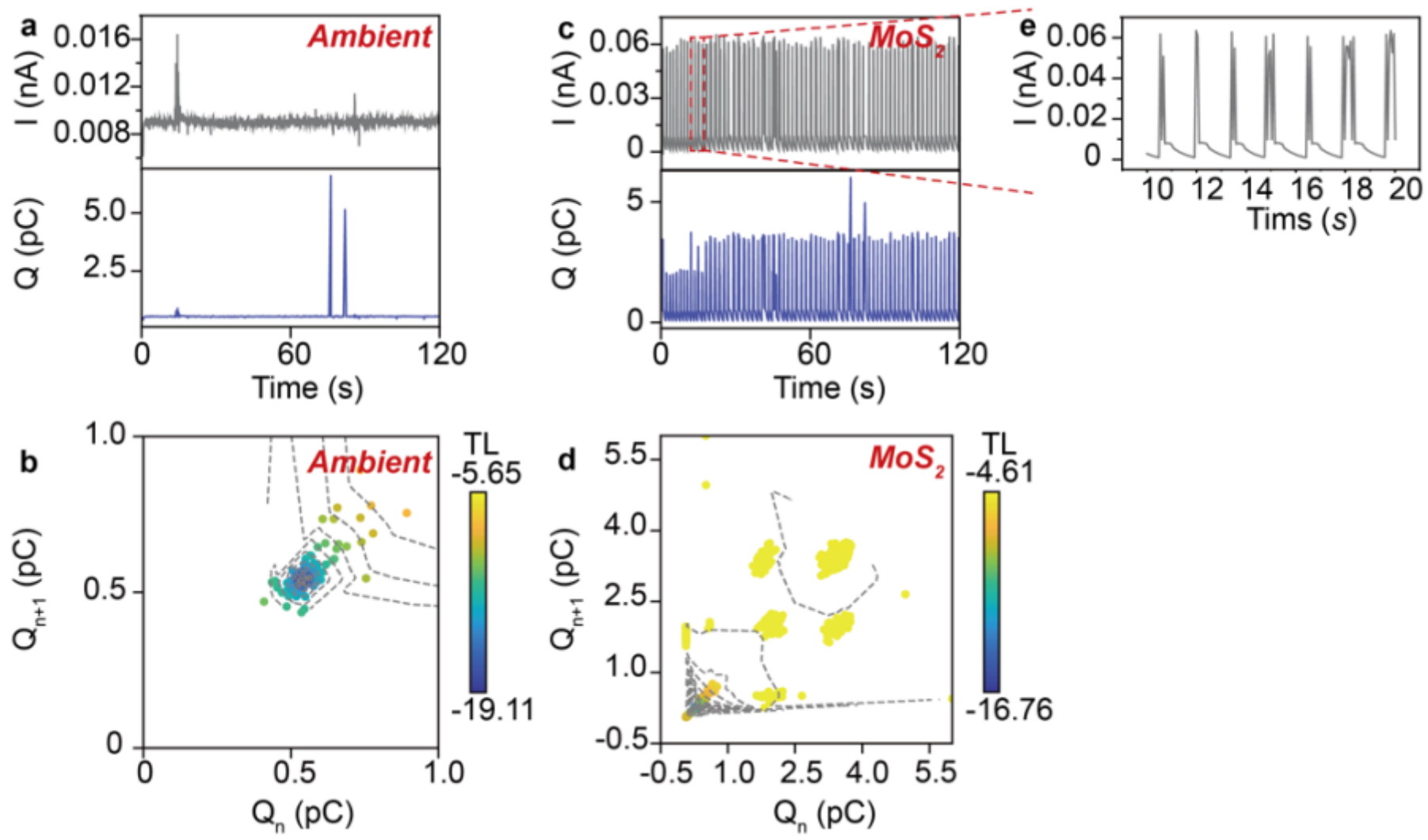

**Figure S10. Control experiments of the ambient noise and a $MoS_2$ device.** Current output and the corresponding cumulative charge fluctuations from (a) ambient noise and (c) a $MoS_2$ device at 0.05 V, 300 K, with (e) showing the detailed current output from the $MoS_2$ device. Ambient noise is measured by suspending the probing electrodes. $MoS_2$ is prepared by liquid-phase exfoliation [4], and is then spin-coated and sandwiched between evaporated top and bottom gold electrodes to fabricate the $MoS_2$ devices. The device fabrication method and the device configuration are the same as that of the 1T' $MoTe_2$ devices. The current output of the $MoS_2$ device shows a distinct random switching behavior, proving random telegraph noise (RTN) [5] from the $MoS_2$ device. The origin of the RTN is suggested to be attributed to charge trapping at the defect sites in the $MoS_2$ material [6,7]. The cumulative charge is integrated during the sampling time interval of 0.067 s. The time-lag plots for the cumulative charge fluctuations of (b) the ambient noise and (d) the $MoS_2$ device: the $\boldsymbol{TL}$ plot of the ambient noise establishes monostable aggregation with weak correlations in the cumulative charge states, meaning weak correlations in the noise states from the ambient noise; the $\boldsymbol{TL}$ plot of the $MoS_2$ device establishes random aggregation regions with strong correlations in the corresponding cumulative charge states, which may be a result of the charge trapping and de-trapping mechanism.

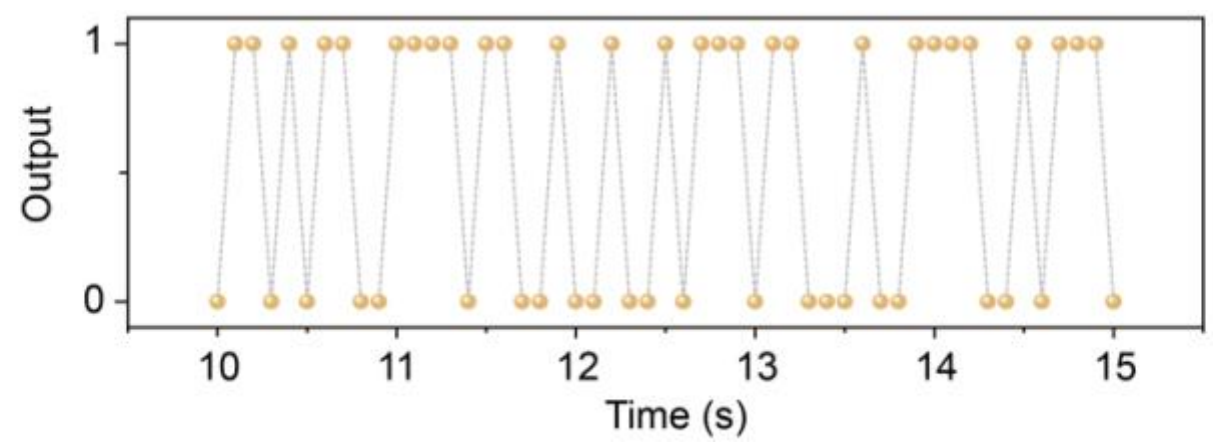

**Figure S11. Random bit string of the 0's and 1's true random numbers generated from "*port 4*" during 10 to 15 seconds** (the full-time-scale true random number output is shown in Fig. 4c). It is shown that the 0's and 1's true random numbers are generated in a random distribution.

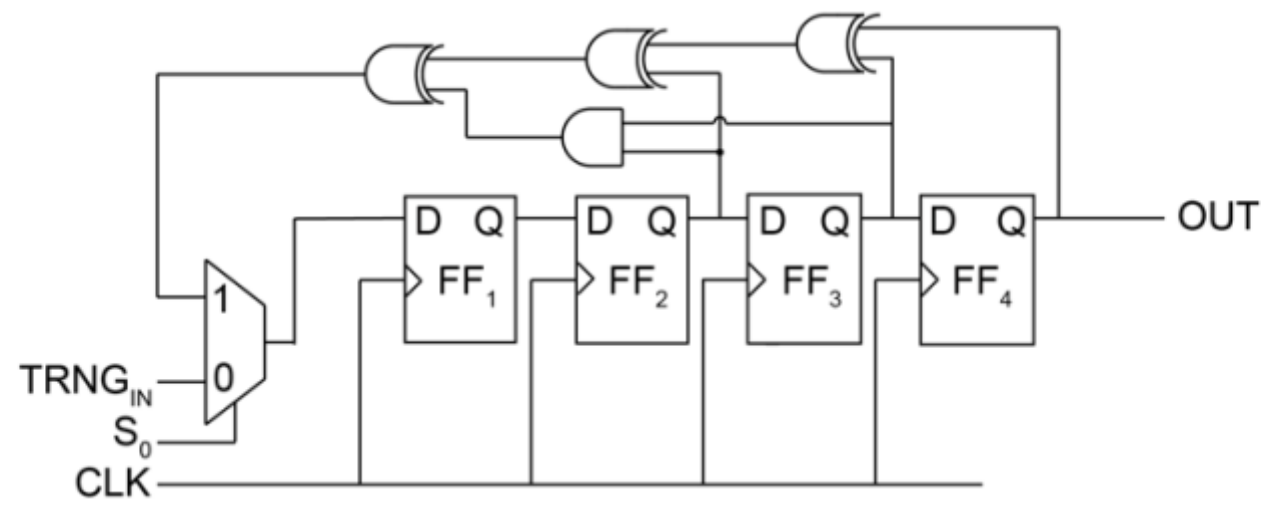

**Figure S12. 4-bit nonlinear feedback shift register (NLFSR) circuit design.** The NLFSR is used to generate high-throughput random numbers from the seed, i.e. the true random numbers, by setting a high clock frequency, e.g. 1 MHz.

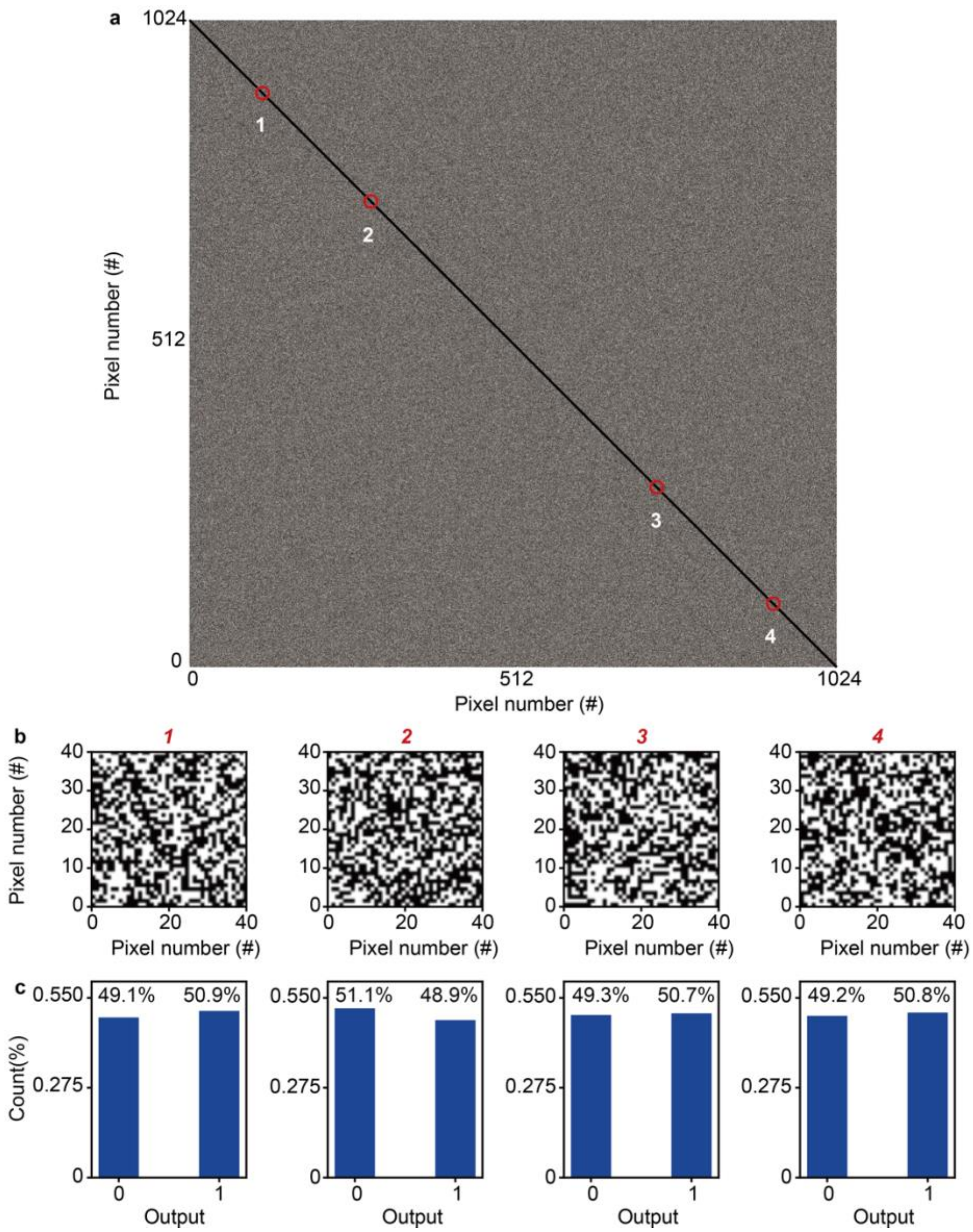

**Figure S13. A 1024*1024 bitmap generated using the high-throughput random numbers.** To plot the bitmap, the string of the high-throughput random numbers in 0's and 1's digits fills the bitmap sequentially from (0, 0) to (1024, 1024). (a) High-data-volume bitmap. (b) Four randomly selected bitmap regions along the diagonal of the high-data-volume bitmap. (c) The corresponding histograms showing the distributions of the 0's and 1's. The ratios of the 0's and 1's are close to 1 to 1.

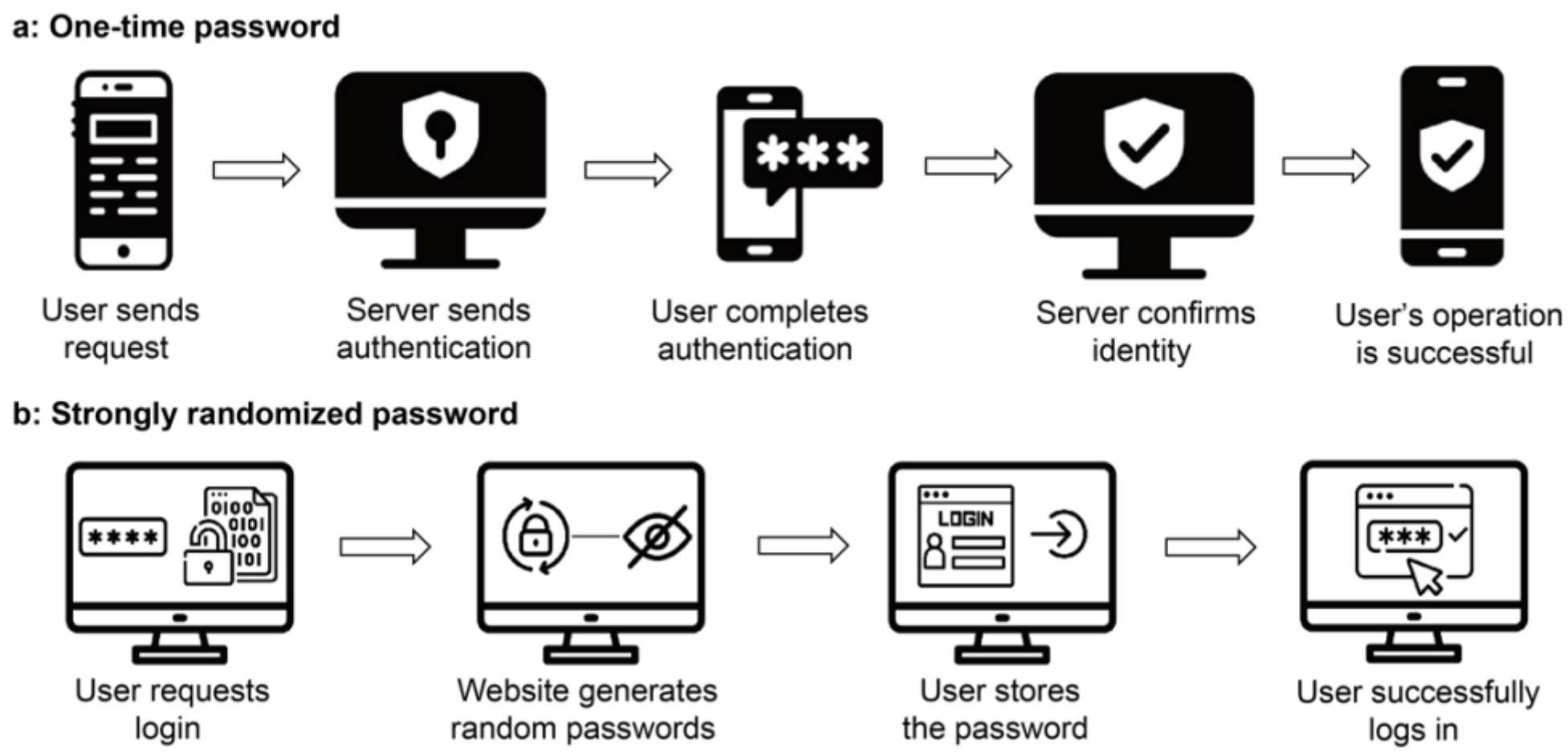

**Figure S14. Password generation.** (a) Schematic generation and the use of one-time password (OTP) [8]. OTP is a unique password code that is valid for only one login/transaction session. It is often used as a second factor in two-factor authentication (2FA) and multi-factor authentication (MFA) systems [8]. (b) Schematic generation and the use of strongly randomized passwords. Strongly randomized passwords can be stored in systems for applications that require a high level of password security.

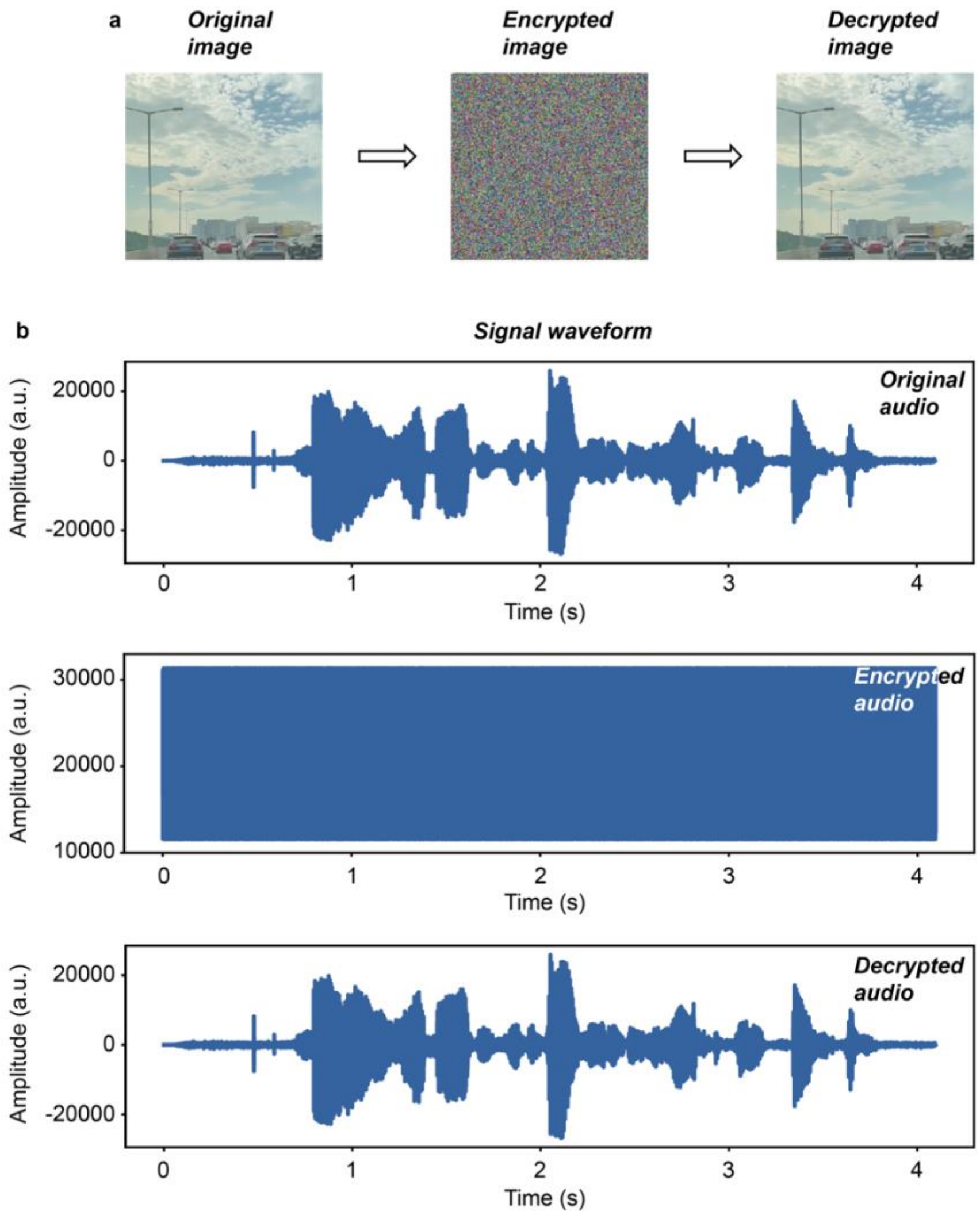

**Figure S15. Data encryption.** (a) Encryption and decryption of an image. The encryption and decryption are conducted by performing the common encryption/decryption operations (e.g. AES encryption/decryption method [9]) on the pixels using the high-throughput random numbers generated. (b) Acoustic spectrograms for audio encryption and decryption, showing a four-second audio, the audio after bit-by-bit encryption, and the audio after decryption.

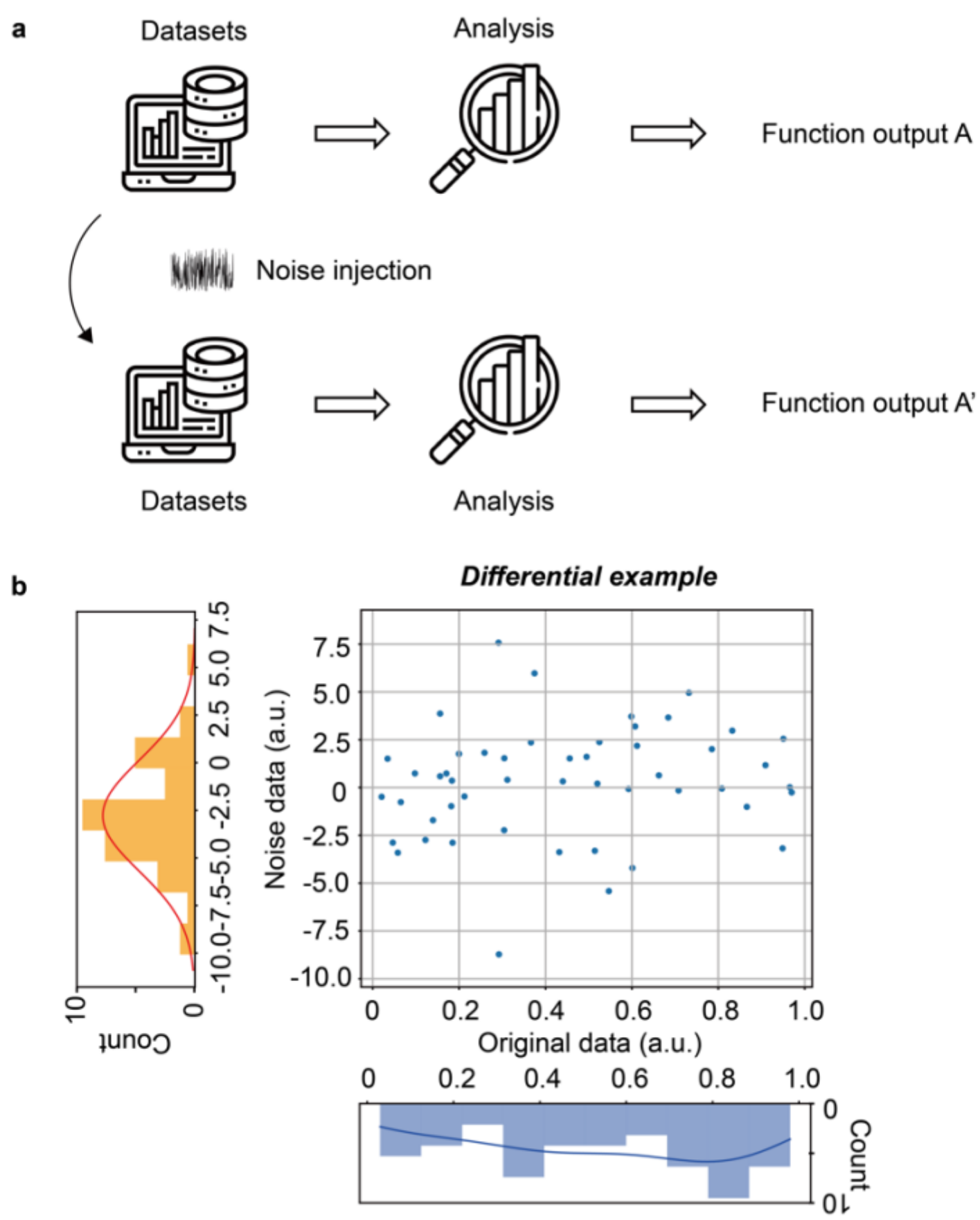

**Figure S16. Differential safeguarding strategy.** (a) Differential data security workflow. Differential data security is a robust framework used in statistical and machine learning analysis of datasets [10]. The core idea is to ensure that the release of the data (or the statistics derived from the data) does not compromise the security of any individuals in the datasets, and that the datasets with the noise perturbation retain the key features. (b) A case example showing injecting noise to a dataset for data security. The left plot shows the noise data, and the bottom plot the original raw data. The noise data is injected into the original data. An example using the differential protection of biometric information such as facial features in neural networks is presented in Fig. 5.

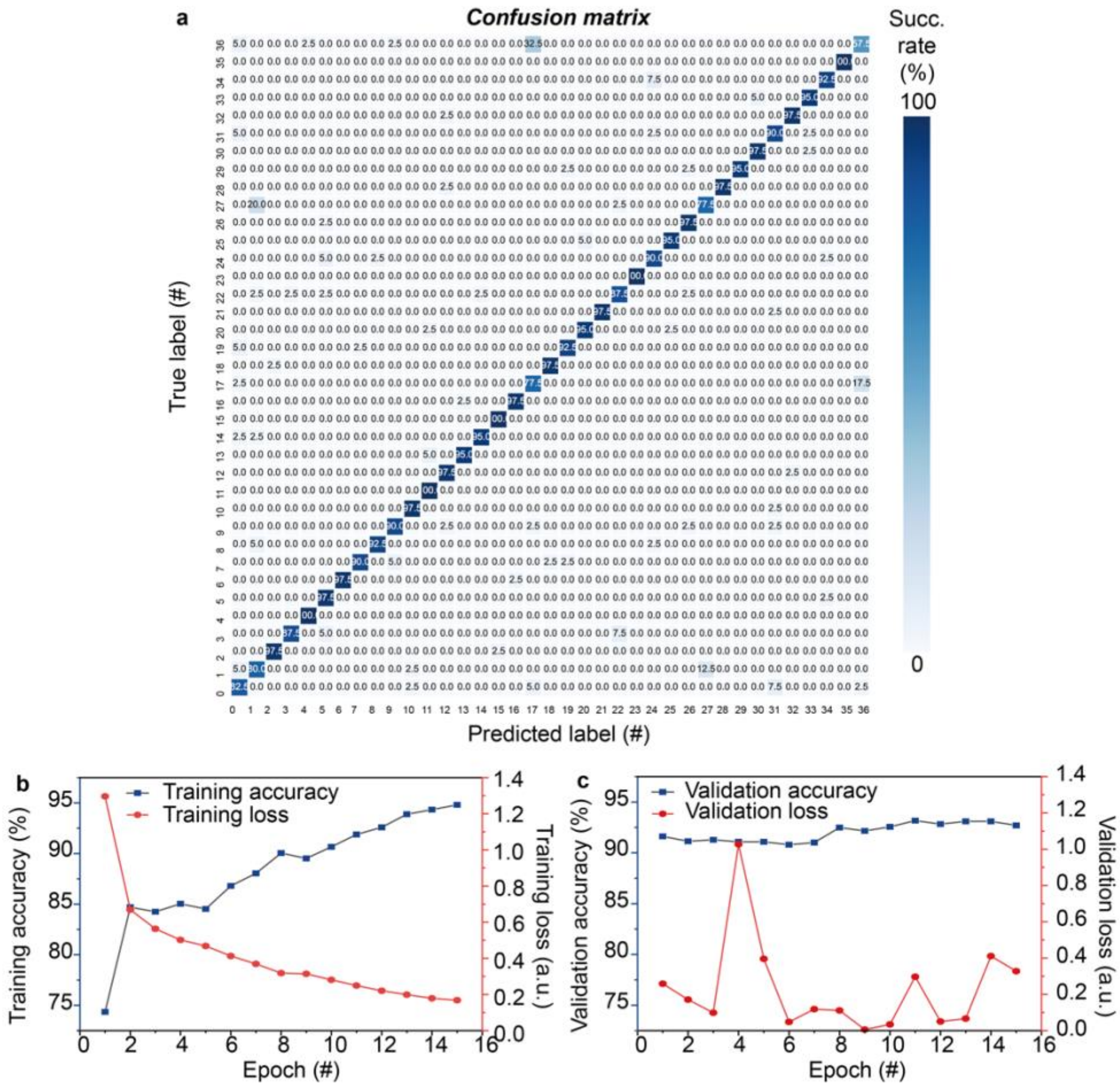

**Figure S17. Confusion matrix without noise perturbation and the training performance of the ResNet model.** (a) Confusion matrix without noise perturbation. $\boldsymbol{x}$ and $\boldsymbol{y}$ coordinates denote the predicated and true labels of the 37 different classifications in the training dataset, i.e. the Oxford pet-iii dataset. The scale (i.e. the success rate of classification) corresponds to the ratio of the number of the correctly predicted labels to the number of the true labels. The values shown in the confusion matrix represent the success rate, showing the trained ResNet variant achieves a good performance (~92%). (b) The training accuracy and training loss of the ResNet variant. (c) The validation accuracy and validation loss of the ResNet variant. Based on the performance of (b) and (c), it can be concluded that the ResNet variant is well-trained for the Oxford pet-iii dataset.

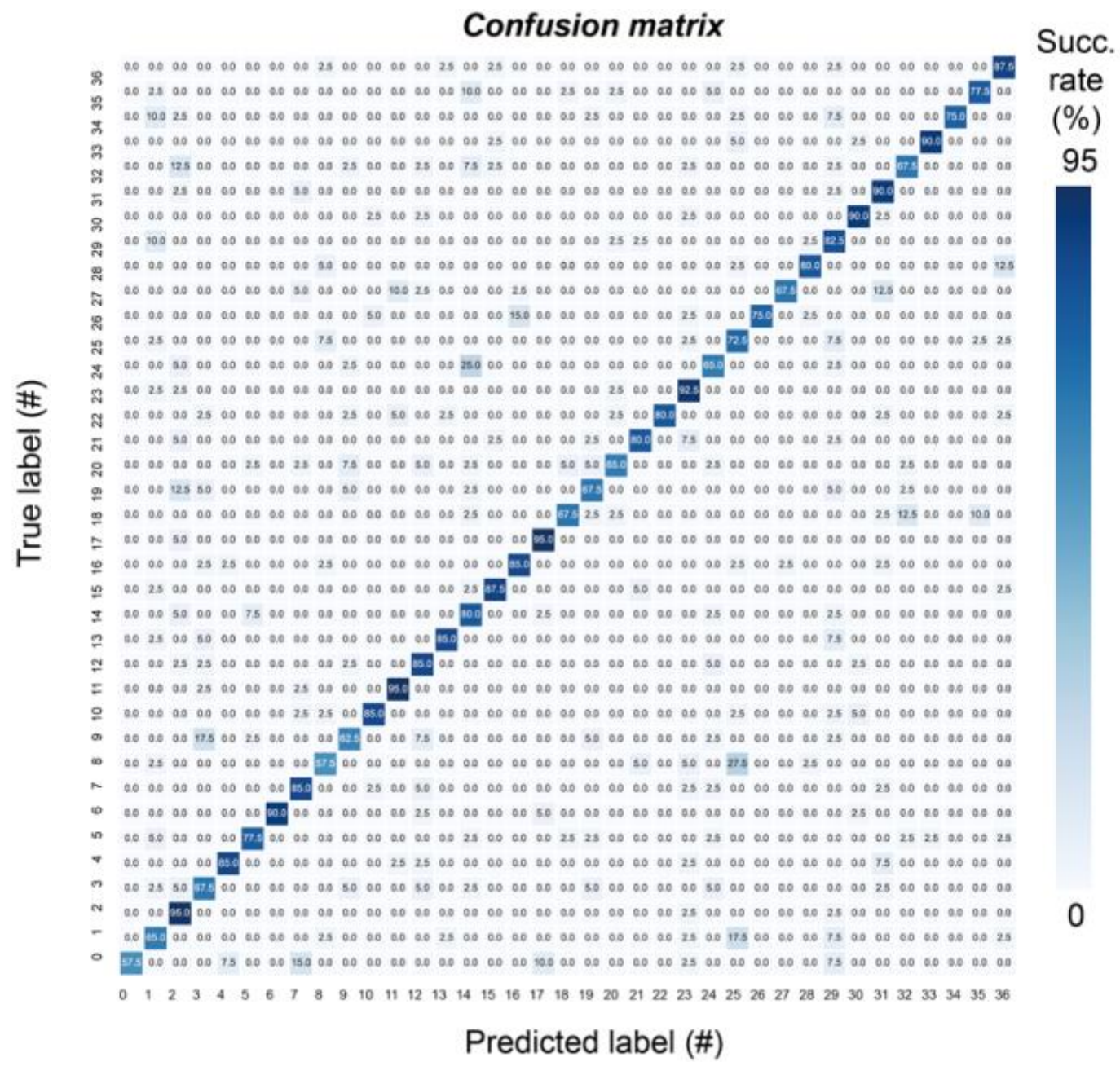

**Figure S18. Confusion matrix after noise perturbation, with the success rate values shown.** $\boldsymbol{x}$ and $\boldsymbol{y}$ coordinates denote the predicated and true labels of the 37 different classifications in the training dataset, i.e. the Oxford pet-iii dataset. The scale (i.e. the success rate of classification) corresponds to the ratio of the number of the correctly predicted labels to the number of the true labels. The values shown in the confusion matrix represent the success rate. Comparison with the confusion matrix in Fig. S17a proves that the noise perturbation reduces the accuracy of the ResNet variant model.

**Supplementary Tables**

**Table S1. NIST test of the true random numbers.**

| | ***P-value*** | ***Success*** | ***Post-processing*** |
|---|---|---|---|
| ***Approximate entropy*** | 1.0 | Success | No |
| ***Block frequency*** | 0.824 | Success | No |
| ***Cumulative sums*** | 0.728, 0.526 | Success | No |
| ***FFT*** | 0.041 | Success | No |
| ***Frequency*** | 0.823 | Success | No |
| ***Linear complexity*** | - | - | Limited throughput |
| ***Longest run*** | 0.445 | Success | No |
| ***Non-overlapping template*** | 0.999 | Success | No |
| ***Overlapping template*** | - | - | Limited throughput |
| ***Random excursions*** | - | Success | No |
| ***Random excursions variant*** | - | Success | No |
| ***Rank*** | - | - | Limited throughput |
| ***Runs*** | 0.017 | Success | No |
| ***Serial*** | 0.499, 0.499 | Success | No |
| ***Universal*** | - | - | Limited throughput |

**Table S2. NIST test of the high-throughput random numbers.**

| | *P-value* | *Proportion* | *Success* | *Post-processing* |
|---|---|---|---|---|
| ***Approximate entropy*** | 0.409 | 1 | Success | No |
| ***Block frequency*** | 0.373 | 1 | Success | No |
| ***Cumulative sums*** | 0.047 | 0.97 | Success | No |
| ***FFT*** | 0.500 | 1 | Success | No |
| ***Frequency*** | 0.057 | 0.98 | Success | No |
| ***Linear complexity*** | 0.371 | 1 | Success | No |
| ***Longest run*** | 0.965 | 1 | Success | No |
| ***Non-overlapping template*** | 0.999 | 1 | Success | No |
| ***Overlapping template*** | 0.873 | 1 | Success | No |
| ***Random excursions*** | 0.525 | 1 | Success | No |
| ***Random excursions variant*** | 0.428 | 1 | Success | No |
| ***Rank*** | 0.696 | 0.99 | Success | No |
| ***Runs*** | 0.630 | 1 | Success | No |
| ***Serial*** | 0.607 | 1 | Success | No |
| ***Universal*** | 0.425 | 1 | Success | No |

**Table S3. Comparison between our random number generator and other reported random number generators.**

| Material and device | Working principle | Circuit complexity | Integration | Exacting difficulty | Power consumption | Temperature window (K) | Applications | Ref. |
|---|---|---|---|---|---|---|---|---|
| Metal/Oxide/Metal memristors | RTN in RRAM | Moderate | high | Moderate | 0.18-18,900 nW | 295.15 | TRNG | 11 |
| Pt/Ag/Ag:$SiO_2$/Pt | Stochastic switching in RRAM | Low | high | Easy | / | 295.15-358.15 | TRNG | 12 |
| Pt/Ti/$NbO_x$/Pt | Stochastic oscillation in RRAM | Moderate | high | Easy | / | 300-390 | TRNG | 13 |
| Silicon-based MTJ | Stochastic switching time in MRAM | High | Moderate | Difficult | 18 pJ/bit | 248-373.15 | TRNG | 14 |
| MTJ | Stochastic magnetization in MRAM | / | Moderate | Difficult | / | 300 | TRNG | 15 |
| MTJ with CNTFET | Stochastic behavior in MRAM | High | Moderate | Difficult | 1.11 pJ/bit | 300 | TRNG | 16 |
| Silicon-based FeFET | Process variations and ferroelectric-material degradation in FeRAM | High | Moderate | Moderate | / | 300 | TRNG | 17 |
| TiN/$HfO_2$/SiON FeFET | Stochastic switching in FeRAM | Low | Moderate | Moderate | / | / | TRNG | 18 |
| Straintronic GFET | Strain-induced reversible cracking | / | / | Easy | / | 298.15 | PUF | 19 |
| $MoS_2$ FET | Inherent stochasticity in the voltage transfer | Low | Moderate | Easy | 30 pJ/bit | 298.15-373.15 | TRNG | 20 |
| $MoS_2$ memtransistor | D2D variation | / | / | Easy | / | 298.15 | PUF | 21 |
| TMO/TMD camouflaged resistors | the properties of TMDs and transition-metal oxides | / | / | Easy | / | / | Hardware obfuscation | 22 |
| $MoS_2$ memtransistor | Programming stochasticity | Low | / | Easy | <1 nJ | / | Stochastic computing | 23 |
| GFET | D2D variation | / | / | Easy | / | 100-380 | PUF | 24 |
| hBN memristor | RTN in RRAM | Moderate | high | Easy | / | 295.15-423.15 | TRNG | 25 |
| **This work** | **Polarization variation** | **Low** | **High** | **Easy** | **50 nW** | **15-300** | **TRNG** | / |

Note the power consumption in this table considers the consumption from the functional device only. (MTJ: magnetic tunnel junction; CNTFET: carbon nanotube field-effect transistor; FeFET: ferroelectric field-effect transistor; GFET: graphene field-effect transistor; FET: field-effect transistor; TMO/TMD: transition metal oxide/transition metal dichalcogenide; RTN: random telegraph noise; RRAM: resistive random-access memory; MRAM: magnetoresistive random-access memory; FeRAM: ferroelectric random-access memory; D2D: device-to-device; TRNG: true random number generator; PUF: physical unclonable function.)

## Supplementary References